\documentclass[journal]{IEEEtran}

\usepackage[caption=false,font=normalsize,labelfont=sf,textfont=sf]{subfig}
\usepackage{amsmath,amsfonts,amssymb,amsthm,amsbsy}
\usepackage[linesnumbered,ruled,vlined]{algorithm2e}
\usepackage{graphicx}
\usepackage{textcomp}
\usepackage{xcolor}
\usepackage{mathtools}
\usepackage[sort,compress]{cite}
\usepackage[bookmarks=true]{hyperref}
\usepackage{enumitem}
\usepackage{dsfont}
\usepackage{breqn}
\usepackage{multirow}
\usepackage{tabularx}

\DeclareMathOperator*{\maximize}{maximize}

\newtheorem{theorem}{Theorem}

\SetKwBlock{Cobegin}{cobegin}{coend}

\def\BibTeX{{\rm B\kern-.05em{\sc i\kern-.025em b}\kern-.08em
    T\kern-.1667em\lower.7ex\hbox{E}\kern-.125emX}}

\begin{document}
	
	\title{5Guard: Isolation-aware End-to-End Slicing of\\5G Networks}

    \author{\IEEEauthorblockN{Mehdi Bolourian, \textit{Graduate Student Member, IEEE}, Noura Limam, \textit{Member, IEEE},\\Mohammad Ali Salahuddin, \textit{Member, IEEE}, Raouf Boutaba, \textit{Fellow, IEEE}\\}\thanks{The authors are with the David R. Cheriton School of Computer Science, University of Waterloo, Waterloo, ON. Email: mbolourian@uwaterloo.ca, noura.limam@uwaterloo.ca, mohammad.salahuddin@uwaterloo.ca, rboutaba@uwaterloo.ca}}
	
	\maketitle
\begin{abstract}
    Network slicing logically partitions the 5G infrastructure to cater to diverse verticals with varying requirements. However, resource sharing exposes the slices to threats and performance degradation, making slice isolation essential. Fully isolating slices is resource-prohibitive, prompting the need for isolation-aware network slicing, where each slice is assigned a tailored isolation level to balance security, usability, and overhead. This paper investigates end-to-end 5G network slicing with resource isolation from the perspective of the infrastructure provider, ensuring compliance with the customers’ service-level agreements. We formulate the online 5G isolation-aware network slicing (5G-INS) as a mixed-integer programming problem, modeling realistic slice isolation levels and integrating slice prioritization. To solve 5G-INS, we propose 5Guard, a novel adaptive framework that leverages an ensemble of custom optimization algorithms to achieve the best solution within resource budget and time constraints. Our results show that 5Guard increases profit by up to 10.1\% in resource-constrained environments and up to 25.4\% in a real-world large-scale network compared to the best-performing individual. Furthermore, we analyze the trade-offs between isolation levels, their impact on resource utilization, and the effects of slice placement, demonstrating significant advantages over baseline approaches that enforce uniform isolation policies.
\end{abstract}

\section{Introduction}\label{intro}
The fifth-generation (5G) mobile networks, empowered by network slicing (NS), promise to support diverse verticals simultaneously \cite{info23, infowk23, tnsm21} on a shared infrastructure. Due to this sharing, these verticals may need security measures \cite{info24} to ensure isolation from vulnerabilities and threats~\cite{ngmn21}, such as distributed denial-of-service (DDoS) or container escape \cite{usenix23}. End-to-end (E2E) isolation is also recognized as a strict demand for sixth-generation (6G) services \cite{ngmn23}. Service customers can request isolated slices, or isolation may be mandated by the security requirements of the slice~\cite{3gpp23}.

Among the most effective approaches to offer the highest security is providing physical isolation of 5G resources across network slices \cite{ngmn22,ngmn21}. However, physical isolation for all network slices can impose significant overhead on the infrastructure provider (InP) and reduce resource utilization.
Thus, it is crucial to offer different levels of isolation~\cite{ngmn22,tmc24} (e.g., physical/complete, logical/semi, or no isolation at all \cite{ngmn22}) to balance security and utilization on a need basis \cite{jlt20,netsoft20,eucnc20,communsur20}. {\color{black} Recent efforts further emphasize the importance of this approach \cite{mobi24, nfv24}.} Consequently, this requires developing new NS schemes that address the above trade-offs in 5G.

A significant challenge in implementing isolation levels within NS lies in ensuring that slices requiring complete isolation are allocated dedicated resources across all segments of the 5G network, which can be hard at scale. One approach to mitigate this challenge is to consolidate resources allocated to other non-fully isolated slices through revisitation, i.e., reusing the same physical entities for realizing these slices, and service migration, which entails transferring services to different physical resources to free up capacity. However, the downtime associated with migration may not be acceptable for all verticals. To address this, slices can be categorized based on priority: high-priority (HP) slices, such as those supporting emergency services, and low-priority (LP) slices, like free public streaming services, and the migration of services enabled in a manner that aligns with the specific requirements and tolerances of each slice.

Another challenge in implementing isolation-aware NS is managing the computational complexity of the mixed-integer resource allocation problem. This complexity can impose scalability and real-world implementation concerns, as exact approaches to tackle this do not often perform well when scaling the network size and the number of slice requests, i.e., the network’s load. Additionally, this complexity can introduce substantial delays, which can be especially intolerable for HP slices. Some algorithms may provide worst-case guarantees for solutions; however, a superior guarantee does not ensure better outcomes in all cases. These complexity or optimality trade-offs can limit their practicality in isolation-aware NS systems that support diverse services. In this regard, an ensemble of custom optimization algorithms can be employed, ranging from fast sub-optimal solutions to slower exact (i.e., optimal) ones. These algorithms can run concurrently by leveraging parallel processing to find the best solution within the resource allocation deadline. This deadline can be considered a quality of service (QoS) requirement, which ensures slices receive their service on time, regardless of the number of slice requests (SR) and the size of the 5G infrastructure network. 

This research aims to answer the following questions while taking into account the outlined issues and challenges:

\begin{enumerate}[leftmargin=*]
    \setlength{\itemindent}{0pt}
    \item What are the trade-offs of incorporating resource isolation levels into E2E 5G NS?
    \item How can we formulate these isolation levels for E2E 5G NS?
    \item What techniques can improve the feasibility and effectiveness of resource isolation in practice?
    \item How can we achieve an efficient isolation-aware NS that ensures timely service provisioning regardless of network load?
\end{enumerate}

To address these research questions, we first formulate an isolation-aware E2E NS problem, modeling three levels of resource isolation while accounting for trade-offs, overheads, and resource usage patterns based on the recommendations. The formulation employs revisitation and migration techniques to ensure feasibility and scalability for complete isolation. To manage the complexity of the problem, we propose an ensemble of algorithms, each with distinct optimality and complexity traits, to achieve an optimal balance within the resource allocation deadline.

Our main contributions are as follows.
\begin{figure*}
	\centering
	\includegraphics[width=0.9\linewidth]{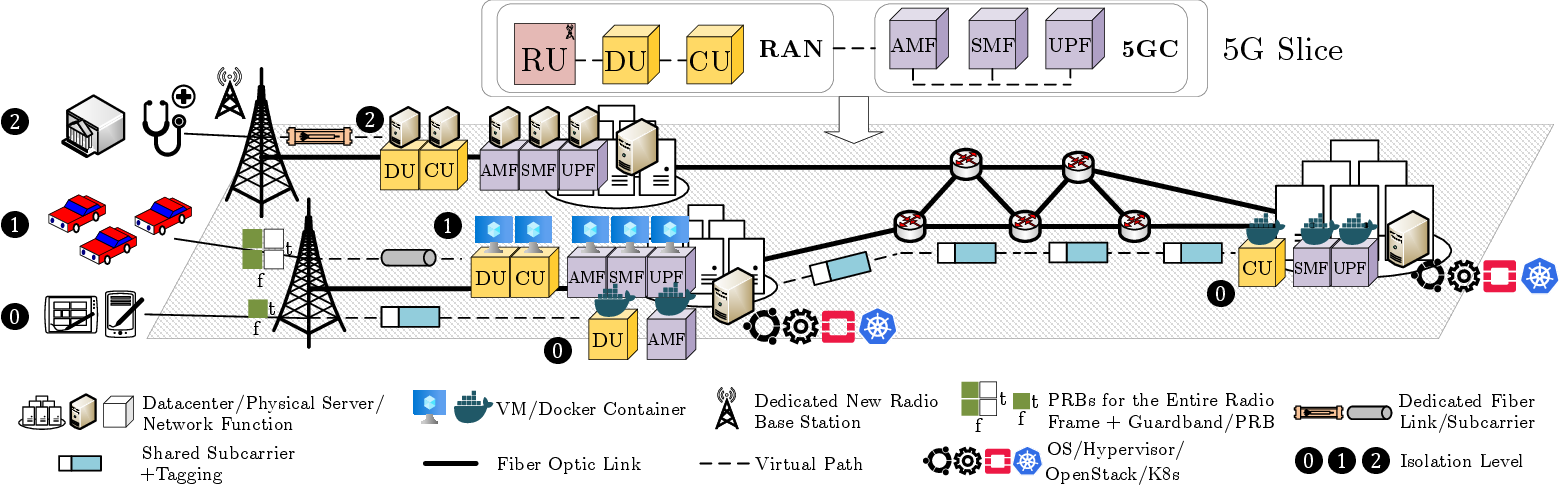}
	\caption{\color{black} The proposed isolation-aware 5G network slicing system model. On top, an E2E 5G slice spanning from the RAN to the 5GC is illustrated, with slice components—NFs, RUs, and VPs—deployable on resources with three isolation levels: L2, L1, and L0. The left-hand side demonstrates three users of 5G slices: L2 for government and medical services, L1 for 5G-V2X services, and L0 for public streaming and entertainment services. Notably, the L2 slice is fully isolated, sharing no resources with other slices. In the example shown, servers' revisitation increases resource efficiency.}\label{system}
\end{figure*}
\begin{itemize}[leftmargin=*]
    \setlength{\itemindent}{0pt}
        \item \textit{\color{black}Problem Formulation:} We mathematically define the 5G isolation-aware NS (5G-INS) problem. It is the first formulation to incorporate E2E resource isolation levels, usage patterns, and overheads. Our online model ensures proper isolation and QoS for SRs, while enhancing slice acceptance through revisitation, service migration, and controlled QoS violations of LP SRs where permissible.
        \item \textit{\color{black}Framework and Algorithm Design:} We propose 5Guard, a novel framework that operates an ensemble of optimization algorithms in parallel at each decision point. These algorithms, designed explicitly for 5G-INS with varying complexity and optimality characteristics, will be selected based on slice priority-imposed deadlines to balance exact and fast solutions.
        \item {\color{black}\textit{Formal Analysis:} We formally analyze the 5G-INS problem, proving its NP-hardness and evaluating the integrality gap. Additionally, we examine the parallel complexity of 5Guard and demonstrate that the relaxed 5G-INS can be effectively expressed as a scalable linear program with polynomial growth in constraints, thus validating 5Guard’s scalability for 5G isolation-aware network slicing.}
        \item \textit{\color{black}Performance Evaluation and Isolation Analysis:} Through extensive experiments, we first show that 5Guard achieves up to {\color{black}10.1\%} higher profit {\color{black}and 25.4\% higher slice admission rate} compared to {\color{black}the best-performing }individual custom-designed 5G-INS algorithms and outperforms the exact algorithm for strict resource allocation deadlines with 100\% admission across all isolation levels. {\color{black}Additionally, we analyze how isolation levels and the placement of slices impact resource efficiency and overhead costs compared to baselines with full isolation and no isolation.}
\end{itemize}

The rest of this paper is organized as follows. Section II presents the slice isolation background and explores the related works. Section III describes the modeling of the 5G network, isolation levels, and slice requests. In Section IV, we formulate the 5G-INS problem {\color{black}and formally analyze its tractability}. We propose the 5Guard framework {\color{black}and study its complexity} in Section V. In Section VI, we numerically evaluate the 5Guard algorithm{\color{black}, and discuss its limitations}. We conclude the paper and instigate future research in Section VII.\newline

\section{Background and Related Works}
{\color{black}In this section, we explore the concept of isolation in 5G NS from different perspectives. We also categorize and review the literature in three related categories: NS, isolation, and isolation-aware NS.}
\subsection{Background}
In cybersecurity, isolation aims to minimize an entity's impact on others, which can serve different purposes \cite{commsur24}. From a security perspective, the goal is to prevent an adversarial entity from initiating attacks on other entities. In addition, effective isolation can mitigate privacy issues \cite{commsur16} or eavesdropping. This isolation dimension is typically called \textit{security isolation} \cite{eucnc20}. Isolating 5G slices can be achieved in various layers, including physical/virtual resources, protocols, applications, and even programming languages. However, not all of these might be relevant or practical for an InP to implement. For instance, the InP does not control the language used in different applications. The layer in which the isolation is enforced can signify its strength.

From the software-defined networking perspective, we can also isolate slices at management and data planes \cite{commsur24, eucnc20}. At the management plane, isolation is enforced through the concept of multi-tenancy. However, even with isolation at higher layers, the security risk can persist if the underlying resources are shared between slices (e.g., container-escape attacks \cite{usenix23}). Therefore, resource isolation is crucial for achieving inter-slice isolation and can span both the data and management planes.

The degree of resource isolation can be measured by comparing the amount of shared resources between slices \cite{eucnc20}. For instance, virtual machines (VM) offer greater isolation than containers due to fewer shared resources (e.g., the operating system (OS)'s kernel). Enforcing isolation or operating in an isolation level can impose additional overheads (e.g., a hypervisor) or resource usage patterns intrinsic to that level and can differ across various 5G resource types, i.e., radio, transport, and mobile-edge computing (MEC)/cloud. For instance, complete (i.e., physical) isolation consumes all the resources of the physical entity, regardless of the resources requested. Studying the impact of resource isolation on 5G NS is one of the main focuses of this work.

\subsection{Related Works}
We explore the related works in three categories as follows.
\subsubsection{State-of-the-Art NS}
Traditional 5G and beyond NS has been studied in the literature. A branch of research in this category abstracted NS as virtual network embedding (VNE) or service function chaining (SFC) \cite{iotj22, compnet24, tnsm233, tnsm232, iotj23, tnsm24, tvt23}. The authors in \cite{iotj22} proposed a VNE solution based on graph neural network (GNN) and reinforcement learning (RL) for dynamic VNE, where both virtual networks (VNs) and physical networks (PNs) are modeled as weighted graphs with changing topologies. Similarly, {\color{black}Moreira et al.} \cite{compnet24} proposed an RL solution for multi-domain and complex infrastructures. In \cite{tnsm233}, the authors proposed a heuristic to solve three formulated problems for multi-cast SFC, readjustment, and expanding to cover the provisioning while considering service migration. However, these approaches fail to adequately capture the impact of isolation due to the overheads and usage patterns across isolation levels, which can significantly impact performance. In addition, most existing RL algorithms require substantial training iterations, and they are unreliable in unseen situations, undermining their practicality for HP SRs. On the other hand, heuristics often demonstrate sub-optimal performance.

\subsubsection{Isolation Concept} In \cite{elcn20}, the authors classified computing isolation methods into language-based, sandbox, VM-based, OS-kernel, and hardware-based techniques. However, a 5G InP might not be able to utilize all these methods. The authors then suggested E2E isolation levels using a weighted power mean function, which does not necessarily yield a meaningful security metric. The authors in \cite{eucnc20} defined isolation levels for 5G resources based on their shared portions. They described isolation approaches for radio, transport, and MEC/cloud based on the latest standards and technologies. However, these studies focused on isolation as a sole concept in 5G and did not evaluate its impact on NS.

\subsubsection{Isolation-Aware NS} A few works studied integrating isolation approaches in NS. The authors in 
\cite{jlt20} proposed a solution for performing isolation-aware 5G RAN slicing enabled by wavelength division multiplexing (WDM). Nevertheless, isolation is enforced at the network domain level. Thus, isolated domains can still be vulnerable depending on their resource isolation level. In \cite{tnsm23}, the authors introduced an RL-based approach to improve slice isolation against attacks. Their solution involved finding the efficient distribution of slices' resources to mitigate the impact of DDoS attacks on a slice from affecting others. Despite overall improvements, their solution does not eliminate the effect for HP SRs and merely considers the servers with small topologies. \color{black}Some works focus on another isolation dimension, namely performance isolation \cite{mobi24, tmc24}. Yarkina et al. in \cite{tmc24} proposed a priority-based performance isolation for 5G radio access network (RAN) and assessed some standard-compliant machine learning (ML) approaches. According to \cite{tmc24}, ML approaches are prone to service requirement violations under high-load conditions. Cheng et al. \cite{mobi24} proposed a slice-aware performance isolation solution for open RAN based on open-source solutions. Although this work can be seen as a proof-of-concept, it does not cover security aspects of isolation or aim to study isolation at levels based on recommendations.

\section{System Model}
{\color{black}In this section, we present our 5G network and isolation models, detailing isolation levels with use cases, and introduce the SR model.}

\subsection{Isolation and Network Model}
{\color{black}As illustrated in Fig. \ref{system},} we consider a 5G infrastructure, consisting of new radio base stations (NR-BS), MEC, and telco cloud servers, as well as single/multi-hop optical links. Without loss of generality, and based on the recommendations in \cite{gsma22,ngmn22}, we define the  following three E2E isolation levels:

\begin{table*}
        \color{black}
        \footnotesize
	\centering
	\caption{Key notations}
	\label{table1}
	\begin{tabular}{ p{0.2cm} p{3.1cm} p{13.3cm}  }
		\hline
		& {\footnotesize \textbf{Symbol}} & {\footnotesize \textbf{Description}}\\
		\hline
		\multirow{5}{*}{\rotatebox{90}{{\footnotesize Variables}}} & $x_{v}^{p}$ & Mapping of NF $v$ to PS $p$\\
            &$y_{w}^{q,\sigma,k}$ & Assigning $k$ PRBs to RU $w$ from frequency slot $\sigma$ of NR-BS $q$\\
            &$z_{vw}^{l,\lambda}$ & Using wavelength slot $\lambda$ for VP $e_{vw}$ from link $l$\\
            &$c_{v}^{p}$ & MIPS capacity of NF $v$ from PS $p$\\
            &$b_{vw}^{pqi,l,\lambda}$ & Bandwidth of VP $e_{vw}$ from PP $e_{pq}^{(i)}$ using wavelength slot $\lambda$ of link $l$\\
		\hline
		\multirow{4}{*}{\rotatebox{90}{{\footnotesize Sets}}} &$\mathcal{R}, \mathcal{R}^{\text{HP}}_{t}, \mathcal{R}^{\text{LP}}_{t}$ & Set of all SRs, excluding pre-embedded HP SRs, and pre-embedded HP and LP SRs at time t\\
		&$\mathcal{V}^{\text{S}}_{r},\mathcal{V}^{\text{R}}_{r},\mathcal{E}_{r}$ & Set of NF and RU nodes and VPs of SR $r$ \\
		&$\mathcal{V}^{\text{P,S}},\mathcal{V}^{\text{P,R}},\mathcal{E}^{\text{P}}$ & Set of PS and NR-BS nodes and PPs\\
            &$\mathcal{L}^{\text{P}},\mathcal{L}_{pqi}$ & Set of all links and links composing PP $e_{pq}^{(i)}$\\
		\hline
		\multirow{10}{*}{\rotatebox{90}{{\footnotesize Parameters}}}&$\gamma_{r}, \kappa_{r}$ & E2E isolation level and sub-level for VNF sharing \\
		&$C_{v}^{\text{REQ}}, K_{w}^{\text{REQ}}, B_{vw}^{\text{REQ}}, D_{vw}^{\text{REQ}}$ & Requested MIPS and PRBs of NF $v$ and RU $w$, and bandwidth and delay of VP $e_{vw}$\\
            &$\Xi_{v}^{\text{REQ}}, \chi_{v,\xi}^{\text{REQ}}$ & Requested number of VNF instances and VNF types of NF $v$\\
		&$R_{r}, \boldsymbol{\rho}_{r}, P^{\min}$ & SR $r$'s revenue and QoS violation cost rate vector and minimum acceptable increased InP cumulative profit from $\mathcal{R}$\\
		&$\Omega, \Phi, \Theta, \beta_{r}$ & MIPS, PRB, and bandwidth cost rate for PSs, RUs, and links, and migration cost rate of SR $r$\\
		&$\Pi^{\max}_{q,\sigma}, B^{\max}_{l,\lambda}$ & Maximum radio and bandwidth capacity of NR-BS $q$ on frequency slot $\sigma$ and link $l$ on wavelength slot $\lambda$\\
            &$\Pi^{\text{PRB}}, N^{\text{FRM}}$ & Radio resources of a PRB and the number of PRBs in a radio frame\\
            &$C_{\text{HHO}}, C_{\text{K8S}},C_{\text{GOS}},\Pi_{\text{FGB}}$ & MIPS overhead of the hypervisor, host OS, and OpenStack, K8s, and guest OS and radio overhead of a guard band\\
            &$C^{\max}_{p}, \Xi_{p,\gamma}^{\max}, D_{pqi}^{\text{RTT}}$ & Maximum MIPS and instances at PS $p$ and isolation level $\gamma$ and round-trip-time delay of PP $e_{pq}^{(i)}$\\
		\hline
	\end{tabular}
\end{table*}

\begin{itemize}[leftmargin=*]
    \setlength{\itemindent}{0pt}
    \item \textbf{Complete isolation (L2)}: This level can support mission-critical (MC) slices, such as emergency, public \cite{3gpp_sa6_cc_apps} and maritime \cite{3gpp232} safety, e-health, government, and military services, as well as virtual operators \cite{3gpp_5gso}. We consider dedicated NR-BSs, fiber links, and physical MEC/cloud server nodes.
    \item \textbf{Semi isolation (L1)}: This level can benefit services that require isolation but cannot afford the high costs of complete isolation, such as ultra-reliable low latency (uRLLC) services and 5G-V2X \cite{3gpp24}. NR-BSs are shared while dedicating physical resource blocks (PRBs) for the entire radio frame to an SR \cite{eucnc20} with additional guard bands, as defined in standards \cite{3gpp242}. For fiber optic links, WDM or elastic optical network (EON) can be used with dedicated wavelength subcarriers and standard guard bands \cite{itu18}. Finally, a VM is instantiated for each network function (NF) instance.
    \item \textbf{Minimum/no isolation (L0)}: This level benefits basic SRs for best-effort 5G services, for instance. PRBs are dedicated to a fraction of the radio frame {\color{black}based on orthogonal frequency-division multiple access (OFDMA)}, as defined in the standards \cite{3gpp242}. Tagging approaches, e.g., multiprotocol label switching (MPLS), can be leveraged to differentiate between slices on the transport while the bandwidth subcarriers are shared. NFs are deployed as containers sharing VMs. We also define the VNF-sharing sub-level, indicating whether containers are shareable. By consolidating SRs in shared entities, i.e., containers, InPs can enhance resource utilization while accommodating more L2 requests.
\end{itemize}

Next, we model the 5G substrate network as an undirected graph with labeled nodes and edges denoted by $\mathcal{G}^{\text{P}}\triangleq\left(\mathcal{V}^{\text{P,R}},\mathcal{V}^{\text{P,S}},\mathcal{E}^{\text{P}}\right)$, where $\mathcal{V}^{\text{P,R}}$ denotes the set of NR-BS nodes, $\mathcal{V}^{\text{P,S}}$ denotes the set of MEC/cloud physical server (PS) nodes, and $\mathcal{E}^{\text{P}}$ denotes the set of physical paths (PPs) between nodes. PPs are composed of a set of interconnected links, denoted by $\mathcal{L}^{\text{P}}$. PS and NR-BS node labels reflect the radio and computing resources' capacity, expressed as a product of a unit of frequency and time and a million instructions per second (MIPS), respectively. PS node labels also represent the maximum number of instances at each isolation level (e.g., VMs or containers). The goal of defining such a limit is to cap the overhead of the hypervisor, guest/host OS, and container orchestration tool (e.g., K8s). PP labels indicate the round-trip-time (RTT) delay and the link bandwidth resources. {\color{black}The key notations are listed in Table \ref{table1}.}

\subsection{Slice Request (SR) Model}
The 5G infrastructure described earlier supports multiple slices with different isolation levels, where each slice has a service-level agreement (SLA). We assume that SRs arrive in an online setting and are stored in an ingress first-in-first-out queue. Based on the SLA, we model an SR, as tuple $r\triangleq\left(\mathcal{V}^{\text{R}}_r, \mathcal{V}^{\text{S}}_r, \mathcal{E}_r, \gamma_r, \kappa_r, R_r, \beta_r, T_r, \boldsymbol{\rho}_r\right)$, where $\mathcal{V}_r^{\text{R}}$, $\mathcal{V}_r^{\text{S}}$, and $\mathcal{E}_r$ denote the set of radio unit (RU) and NF nodes and virtual paths (VPs) of the VN, respectively. $\gamma_r$ and $\kappa_r$ denote the isolation level and L0 VNF sharing sub-level, respectively. $R_r$ denotes the revenue, $\beta_r$ denotes the cost of a unit of migration, $T_r$ denotes the resource allocation deadline, $\boldsymbol{\rho}_r$ is the SR's cost vector of a unit of QoS violation for radio, MIPS, and bandwidth resources. For recurring slices, the customer must request a new slice at the end of the slice's lifetime.

\section{5G-INS  Formulation}
{\color{black}In this section, we formally define 5G-INS as the 5G isolation-aware network slicing problem, formulating it through constraints and an objective function. We also analyze its complexity and reduce it to a manageable form.}
\subsection{Problem}
We explore the 5G-INS problem from the InP's perspective{\color{black}, responsible for making slicing decisions to ensure customers receive services aligned with their SLAs. }The goal is to embed each incoming SR, denoted by $r_t$, upon arrival at time step $t$, and reconfigure the pre-embedded \textcolor{black}{LP} slices in the set denoted by \textcolor{black}{$\mathcal{R}^{\text{LP}}_t$} as needed. This should consider the slices' priority and associated costs to meet isolation level $\gamma_r$ and the QoS requirements for all SRs. We assume the InP aims to maximize profit and must determine how to map VNs best and allocate resources on PN $\mathcal{G}^{\text{P}}$. In addition, InP should ensure that it does not surpass the available resources while factoring in the overheads and resource usage patterns of isolation levels. {\color{black}Considering these constraints, we assume that InP intends to admit all SRs arriving at different time steps, unless infeasible.}

\subsection{Decision Variables}
Let $\mathcal{R}\triangleq{\color{black}\mathcal{R}^{\text{LP}}_t}\cup\{r_t\}$ denote the set of all SRs{\color{black}, except the pre-embedded HP SRs,} at time step $t$. For mapping NFs to PSs, we define binary variable $x_v^p$ in vector $\boldsymbol{x}$ that indicates whether NF $v\in\mathcal{V}_r$ of SR $r\in\mathcal{R}$ is mapped to PS $p\in\mathcal{V}^{\text{P}}$. We also denote $c_v^p\in \mathbb{R}^+$ in vector $\boldsymbol{c}$ as the amount of MIPS allocated to NF $v\in\mathcal{V}_r$ of SR $r\in\mathcal{R}$ from PS $p\in\mathcal{V}^{\text{P}}$. Since radio resources are typically allocated in PRBs, they are inherently integer. Additionally, resources are released at the end of a slice's lifetime. Thus, {\color{black}considering pre-embedded HP SRs' allocated resources}, they can be fragmented into a set of frequency slots denoted by $\mathcal{S}_q$ for NR-BS $q\in\mathcal{V}^{\text{P,R}}$. Thus, we define binary variable $y_w^{q,\sigma,k}$ in vector $\boldsymbol{y}$ indicating whether $k$ PRBs are allocated to RU $w\in\mathcal{V}_r^{\text{R}}$ of SR $r\in\mathcal{R}$ from the frequency slot $\sigma\in\mathcal{S}_q$ of NR-BS $q\in\mathcal{V}^{\text{P,R}}$.

Bandwidth fragmentation can similarly occur for fiber optic links. Thus, we denote $\mathcal{D}_l$ as the set of wavelength slots for link $l\in\mathcal{L}^{\text{P}}$. We define binary variable $z_{vw}^{l,\lambda}$ in vector $\boldsymbol{z}$ that indicates whether VP $e_{vw}\in\mathcal{E}_r$, the VP connecting NF/RU nodes $v\in\mathcal{V}_r^{\text{S}}\cup\mathcal{V}_r^{\text{R}}$ and $w\in\mathcal{V}_r^{\text{S}}\cup\mathcal{V}_r^{\text{R}}$ of SR $r\in\mathcal{R}$, uses the wavelength slot $\lambda\in\mathcal{D}_l$ of link $l\in\mathcal{L}^{\text{P}}$. To avoid bottlenecks and increase utilization, we allow bandwidth splitting between the links of multiple PPs. Thus, we define variable $b_{vw}^{pqi,l,\lambda}\in \mathbb{R}^+$ in vector $\boldsymbol{b}$ as the bandwidth allocated to VP $e_{vw}\in\mathcal{E}_r$ of SR $r\in\mathcal{R}$ from  slot $\lambda\in\mathcal{D}_l$ of link $l\in\mathcal{L}_{pqi}$, where $\mathcal{L}_{pqi}$ denotes the set of links on PP $e_{pq}^{(i)}\in\mathcal{E}^{\text{P}}$.
\subsection{NF and RU Mapping Constraints}
\subsubsection{\textcolor{black}{Mapping Uniqueness}} To ensure each NF and RU is precisely mapped to a single PS and NR-BS, respectively, we enforce,
\begin{equation}\label{eq1}
    \sum_{p\in\mathcal{V}^{\text{P,S}}}  x_v^p  =  1,\quad\forall v\in \mathcal{V}^{\text{S}}_{r}, r\in\mathcal{R},
\end{equation}
\begin{equation}\label{eq2}
    \sum_{q\in\mathcal{V}^{\text{P,R}}}  \sum_{\sigma\in\mathcal{S}_q}  \sum_{k=1}^{K_w^{\text{REQ}}}  y_w^{q,\sigma,k}  =  1,\quad\forall w\in \mathcal{V}^{\text{R}}_{r},  r\in\mathcal{R}
\end{equation}
where $K_w^{\text{REQ}}$ denotes the number of PRBs requested by RU $w\in\mathcal{V}_r^{\text{R}}$ of SR $r\in\mathcal{R}$. For RUs, a single frequency slot can be selected in the mapped NR-BS to maintain spectral contiguity. The summation over $k$  prevents selecting multiple values for the assigned PRBs. {\color{black}Note that for pre-embedded HP SRs, we reuse the previous NF and RU mappings, since migration can lead to service degradation.}

\subsubsection{\textcolor{black}{Complete Isolation Mapping}} \textcolor{black}{For ease of exposition, we first define $f_{\hat{p}}(\mathcal{\hat{V}}, \mathcal{\hat{R}}, \boldsymbol{a}^{\hat{p}})$ as the function that checks how many L2 SRs have at least one virtual entity, i.e., NF, RU, or VP, in set $\mathcal{{\hat{V}}}\triangleq\bigcup_{r\in\mathcal{\hat{R}}}^{\gamma_r=2} \mathcal{\hat{V}}_r$ that are mapped to physical entity $\hat{p}$, i.e., a PS, NR-BS, or PP, based on the mapping decision vector $\boldsymbol{a}^{\hat{p}}\triangleq(a_{\hat{v}}^{\hat{p}})_{\hat{v}\in\mathcal{\hat{V}}}$. Thus,
\begin{equation}\label{eq3}
    f_{\hat{p}}(\mathcal{\hat{V}}, \mathcal{\hat{R}}, \boldsymbol{a}^{\hat{p}})  \triangleq   \sum_{\substack{r\in\mathcal{\hat{R}}}}^{\gamma_r=2}  \max_{\hat{v}\in\mathcal{\hat{V}}_r}  a_{\hat{v}}^{\hat{p}}
\end{equation}
}
For L2, a PS or NR-BS must be mapped to at most one NF or RU, respectively. \textcolor{black}{Accordingly, we need to enforce the following L2 constraints for each PS and NR-BS, respectively.
\begin{equation}\label{eq4}
    f_{p}(\mathcal{V}^{\text{S}}, \mathcal{R}, \boldsymbol{x}^{p})  \leq  1,\quad\forall p\in\mathcal{V}^{\text{P,S}}
\end{equation}
where $\boldsymbol{x}^{p}\triangleq(x_v^p)_{v\in\mathcal{V}_r^{\text{S}}}^{r\in\mathcal{R}}$ and $\mathcal{V}^{\text{S}}\triangleq\bigcup_{r\in\mathcal{R}}\mathcal{V}^{\text{S}}_r$.
\begin{equation}\label{eq5}
    f_{q}(\mathcal{V}^{\text{R}}, \mathcal{R}, \boldsymbol{y}^{q})  \leq  1,\quad\forall q\in\mathcal{V}^{\text{P,R}}
\end{equation}
where $\boldsymbol{y}^{q}  \triangleq  (\sum_{\sigma\in\mathcal{S}_q}\sum_{k=1}^{K_w^{\text{REQ}}}y_{w}^{q,\sigma,k})_{w\in\mathcal{V}_r^{\text{R}}}^{r\in\mathcal{R}}$ and $\mathcal{V}^{\text{R}}  \triangleq  \bigcup_{r\in\mathcal{R}}\mathcal{V}^{\text{R}}_r$.
}

To prevent other slices from being mapped to the L2 slices' PSs and NR-BSs,
\textcolor{black}{
\begin{equation}\label{eq6}
		x_v^p  \leq  1-f_p(\mathcal{V}^{\text{S}}, \mathcal{R}\backslash\{r\},\boldsymbol{x}^{p}),\quad\forall v\in\mathcal{V}^{\text{S}}_{r}, r\in\mathcal{R},  p\in\mathcal{V}^{\text{P,S}}
\end{equation}
\begin{equation}\label{eq7}
	\begin{aligned}
		\sum_{\sigma\in\mathcal{S}_q}  \sum_{k=1}^{K_w^{\text{REQ}}}  y_{w}^{q,\sigma,k}  \leq  1 - f_q(\mathcal{V}^{\text{R}}, \mathcal{R}\backslash\{r\}, \boldsymbol{y}^{q}),&\\
        \forall w\in\mathcal{V}^{\text{R}}_{r}, r\in\mathcal{R},  q\in\mathcal{V}^{\text{P,R}}&
	\end{aligned}
\end{equation}}

\subsubsection{\textcolor{black}{Coordination and Overprovisioning Prevention}} When an NF is not mapped to a PS, the allocated capacity associated with that mapping should also be zero, and vice versa. Additionally, to prevent overprovisioning,
\begin{equation}\label{eq8}
	\delta x_v^p\leq c_v^p\leq C^{\text{REQ}}_{v}x_v^p,\quad\forall p\in\mathcal{V}^{\text{P,S}}, v\in\mathcal{V}^{\text{S}}_{r},  r\in\mathcal{R}
\end{equation}
where $C_v^{\text{REQ}}$ and $\delta$ are MIPS requested by NF $v\in\mathcal{V}_r^{\text{S}}$ of SR $r\in\mathcal{R}$ and the smallest MIPS capacity that can be allocated on PSs, respectively. Similarly, for radio resources,
\begin{equation}\label{eq9}
	\sum_{\sigma\in\mathcal{S}_q}\sum_{k=1}^{K_w^{\text{REQ}}}ky_{w}^{q,\sigma,k}\leq K_w^{\text{REQ}},\quad\forall q\in\mathcal{V}^{\text{P,R}},w\in\mathcal{V}^{\text{R}}_{r},  r\in\mathcal{R}
\end{equation}
\subsubsection{\textcolor{black}{Maximum Capacity}} The allocated MIPS to each PS across all isolation levels should not exceed its capacity. As mentioned earlier, isolation levels contribute to specific MIPS overheads, either from the isolation itself or are required for operation at that isolation level. These overheads arise from {\color{black}OpenStack,} a host OS, hypervisor, and several guest OSs and container orchestration tools, i.e., K8s, which scale with the number of VMs.
\begin{figure}
	\centering
	\includegraphics[width=0.9\linewidth]{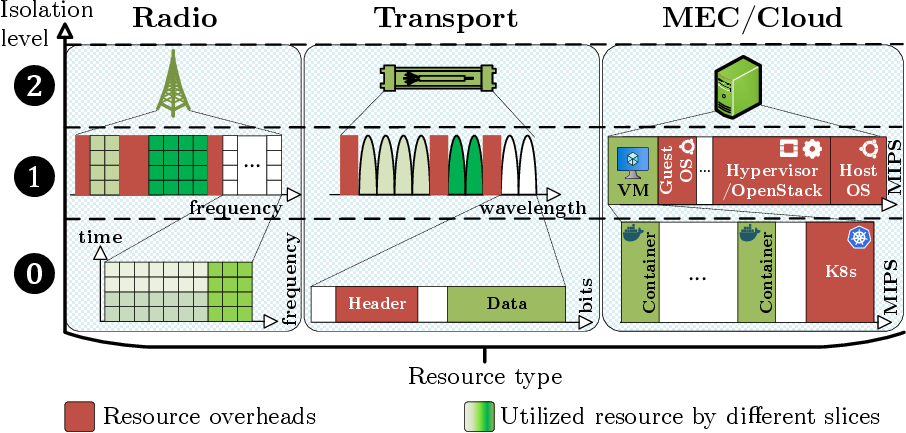}
	\caption{\textcolor{black}{The proposed isolation levels and resource overheads: For radio resources, these include dedicated NR-BS, PRBs for the entire radio frame with guard bands, and PRBs allocated via OFDMA. Transport isolation uses dedicated fiber, subcarriers with WDM/EON, and shared subcarriers with tagging. MEC/cloud employs servers, VMs, and containerized K8s pods.
    }}\label{isolation}
\end{figure}

\textcolor{black}{To calculate these, we first need to derive the number of VMs for L0 SRs. According to Fig. \ref{isolation}, this controls how much overhead is imposed by the container orchestration tool (e.g., K8s) and guest OSs. This paper assumes containers are placed in VMs to improve isolation from L1. This is also common in creating K8s clusters, the substrate to run the containers. For ease of exposition, we first define $h_p^{\text{L0,0,U}}$, $h_p^{\text{L0,0,S}}$ as the number of VMs of L0 SRs using PS $p\in\mathcal{V}^{\text{P,S}}$ with isolation sub-level of 0 for two types of VNFs: 1) unsharable and 2) sharable, respectively as follows,
\begin{equation}\label{eq10}
	\begin{aligned}
    h_p^{\text{L0,0,U}}\triangleq\sum_{\substack{r\in\mathcal{R}\cup\mathcal{R}^{\text{HP}}_t}}^{ \substack{\gamma_r=\kappa_r=0}}\sum_{\substack{v\in\mathcal{V}^{\text{S}}_r}}\sum_{\substack{\xi=1 \\ \chi_{v,\xi}^{\text{REQ}}=\chi^{\text{U}}}}^{\Xi_v^{\text{REQ}}}\frac{\tilde{x}_v^p}{\Xi_{p,0}^{\max}}
	\end{aligned}
\end{equation}
\begin{equation}\label{eq11}
    h_p^{\text{L0,0,S}}\triangleq\sum_{\chi=0}^{\chi_{\max}^{\text{S}}}\max_{\substack{v\in\mathcal{V}_r^{\text{S}},  1\leq\xi\leq\Xi_v^{\text{REQ}}, \\ \gamma_r=\kappa_r= 0, r\in\mathcal{R}\cup\mathcal{R}^{\text{HP}}_t}}\frac{\mathds{1}_{\{\chi_{v,\xi}^{\text{REQ}}=\chi\}}\tilde{x}_v^p}{\Xi_{p,0}^{\max}}
\end{equation}}
\textcolor{black}{where
\[
\tilde{x}_v^p \triangleq 
\begin{cases} 
x_v^p, & \forall v \in \mathcal{V}_r^{\text{S}}, r \in \mathcal{R} \\
X_v^p, & \forall v \in \mathcal{V}_r^{\text{S}}, r \in \mathcal{R}^{\text{HP}}_t
\end{cases}
\] with $X_v^p$ denoting the existing NF mapping values for pre-embedded HP SRs. In addition,  $\mathds{1}_{\{\cdot\}}$ is the indicator function, and $\Xi_{p,0}^{\max}$ represents the maximum number of containers recommended by the orchestration tool to avoid resource contention. Here, $\chi_{v,\xi}^{\text{REQ}}$ denotes the type of VNF instance $\xi\in[1,\Xi_v^{\text{REQ}}]$ of NF $v\in\mathcal{V}_r^{\text{S}}$ requested by SR $r\in\mathcal{R}$, with $\Xi_v^{\text{REQ}}$ denoting the number of requested VNF instances. In addition, $\chi^{\text{U}}$ and $\chi_{\max}^{\text{S}}$ denote the unsharable VNF type and the maximum number of sharable types, respectively. The $\max$ function used in (\ref{eq11}) ensures that one container is instantiated for an individual sharable type at most. Then, by applying a division by $\Xi_{p,0}^{\max}$, the number of VMs is derived.}

\textcolor{black}{Similarly, we define $h_p^{\text{L0,1}}$ as the number of VMs of L0 SRs with isolation sub-level of 1, where no VNF sharing is allowed, as follows.
\begin{equation}\label{eq12}
	\begin{aligned}
    h_p^{\text{L0,1}}\triangleq\sum_{r\in\mathcal{R}\cup\mathcal{R}^{\text{HP}}_t}^{\substack{\gamma_r=0, \kappa_r=1}}\sum_{\substack{v\in\mathcal{V}^{\text{S}}_r}}\frac{\Xi_v^{\text{REQ}}\tilde{x}_v^p}{\Xi_{p,0}^{\max}},\\
	\end{aligned}
\end{equation}
Here, a container is instantiated for each requested instance, similar to (\ref{eq10}).}

\textcolor{black}{Accordingly, we can derive the overall number of VMs for L0 SRs, denoted by $h_p^{\text{L0}}$, as follows.
\begin{equation}\label{eq13}
	\begin{aligned}
    h_p^{\text{L0}}=\left\lceil h_p^{\text{L0,0,U}}+h_p^{\text{L0,0,S}}+h_p^{\text{L0,1}}\right\rceil,\\
	\end{aligned}
\end{equation}
where $\lceil\cdot\rceil$ avoids deriving a fractional number.}

\textcolor{black}{By utilizing (\ref{eq13}), we can derive the total MIPS overhead of deploying the isolation levels for PS $p\in\mathcal{V}^{\text{P,S}}$, denoted by $C_{p,\text{TOT}}^{\text{OVH}}$, as follows.
\begin{equation}\label{eq14}
	\begin{aligned}
    C_{p,\text{TOT}}^{\text{OVH}}=\left(C_{\text{K8S}}+C_{\text{GOS}}\right)h_p^{\text{L0}}+C_{\text{HHO}}\max_{\substack{v\in\mathcal{V}^{\text{S}}_{r}, \gamma_r\leq 1, \\ r\in\mathcal{R}\cup\mathcal{R}^{\text{HP}}_t}}\tilde{x}_v^p&\\
        +C_{\text{GOS}}\sum_{\substack{r\in\mathcal{R}\cup\mathcal{R}^{\text{HP}}_t}}^{\gamma_r = 1}\sum_{\substack{v\in\mathcal{V}^{\text{S}}_r}}\Xi_v^{\text{REQ}}\tilde{x}_v^p&
	\end{aligned}
\end{equation}
where $C_{\text{K8S}}$, $C_{\text{GOS}}$, and $C_{\text{HHO}}$ are the maximum MIPS overhead of the container orchestration tool (i.e., K8s), guest OS, and hypervisor, host OS, and OpenStack, respectively.}

\textcolor{black}{Now that we have derived the total MIPS overhead, by utilizing (\ref{eq14}), the constraint can be determined as follows.}
\begin{equation}\label{eq15}
	\begin{aligned}
    \sum_{r\in\mathcal{R}\cup\mathcal{R}^{\text{HP}}_t}\sum_{\substack{v\in\mathcal{V}^{\text{S}}_{r}}}\tilde{c}_v^p\leq C^{\max}_{p}-C_{p,\text{TOT}}^{\text{OVH}},\quad\forall p\in\mathcal{V}^{\text{P,S}}&
	\end{aligned}
\end{equation}
{\color{black}where \[
\tilde{c}_v^p \triangleq 
\begin{cases} 
c_v^p, & \forall v \in \mathcal{V}_r^{\text{S}}, r \in \mathcal{R} \\
C_v^p, & \forall v \in \mathcal{V}_r^{\text{S}}, r \in \mathcal{R}^{\text{HP}}_t
\end{cases}
\] with $C_v^p$ denoting the existing allocated MIPS capacity to $v\in\mathcal{V}_r^{\text{S}}$ for pre-embedded HP SR $r\in\mathcal{R}^{\text{HP}}_t$ and} $C_p^{\max}$ is the MIPS capacity of PS $p\in\mathcal{V}^{\text{P,S}}$.

Considering the VMs instantiated for slices at L1 and for hosting containers at L0, there should be a limit on the number of VMs on PS $p\in\mathcal{V}^{\text{P,S}}$, denoted $\Xi_{p,1}^{\max}$. Otherwise, the MIPS capacity overhead of the hypervisor can grow significantly by constantly adding VMs. Therefore,
\begin{equation}\label{eq16}
    \begin{aligned}
h_p^{\text{L0}}+\sum_{\substack{r\in\mathcal{R}\cup\mathcal{R}^{\text{HP}}_t}}^{\gamma_r=1}\sum_{\substack{v\in\mathcal{V}^{\text{S}}_{r}}}\Xi_v^{\text{REQ}}\tilde{x}_v^p\leq \Xi_{p,1}^{\max},\quad\forall p\in\mathcal{V}^{\text{P,S}}&
    \end{aligned}
\end{equation}

For each NR-BS, we should not surpass its maximum radio resources at each frequency slot, considering the coexistence of L0 and L1 and the frequency guard bands' overhead. {\color{black}First, let $\Pi_{\text{FGB}}^{\text{OVH}}$ denote the radio overhead imposed by each frequency guard band at L1. We can then derive the total utilized radio and radio overhead for frequency slot $\sigma\in\mathcal{S}_q$ of NR-BS $q\in\mathcal{V}^{\text{P,R}}$, denoted by $\Pi_{q,\text{TOT}}^{\sigma,\text{UTL}}$ and $\Pi_{q,\text{TOT}}^{\sigma,\text{OVH}}$, respectively, as follows.
\begin{equation}\label{eq17}
    \begin{aligned}
    \Pi_{q,\text{TOT}}^{\sigma,\text{UTL}}=&\Pi^{\text{PRB}}\Bigg(\sum_{\substack{r\in\mathcal{R}\cup\mathcal{R}^{\text{HP}}_t}}^{\gamma_r\neq 1}\sum_{\substack{w\in\mathcal{V}^{\text{R}}_{r}}}\sum_{k=1}^{K_w^{\text{REQ}}} k\tilde{y}_{w}^{q,\sigma,k}+\\
    &N^{\text{FRM}}\sum_{\substack{r\in\mathcal{R}\cup\mathcal{R}^{\text{HP}}_t}}^{\gamma_r=1}\sum_{\substack{w\in\mathcal{V}^{\text{R}}_{r}}}\sum_{k=1}^{K_w^{\text{REQ}}} \left\lceil\frac{k\tilde{y}_{w}^{q,\sigma,k}}{N^{\text{FRM}}}\right\rceil\Bigg)
    \end{aligned}
\end{equation}
\begin{equation}\label{eq18}
    \Pi_{q,\text{TOT}}^{\sigma,\text{OVH}}=\sum_{\substack{r\in\mathcal{R}\cup\mathcal{R}^{\text{HP}}_t}}^{ \gamma_r=1}\sum_{\substack{w\in\mathcal{V}^{\text{R}}_{r}}}\sum_{k=1}^{K_w^{\text{REQ}}}\tilde{y}_{w}^{q,\sigma,k}\Pi_{\text{FGB}}^{\text{OVH}}
\end{equation}
where \[
\tilde{y}_{w}^{q,\sigma,k} \triangleq 
\begin{cases} 
y_{w}^{q,\sigma,k}, & \forall w \in \mathcal{V}^{\text{R}}_r, r \in \mathcal{R} \\
Y_{w}^{q,\sigma,k}, & \forall w \in \mathcal{V}^{\text{R}}_r, r \in \mathcal{R}^{\text{HP}}_t
\end{cases}
\] with $Y_{w}^{q,\sigma,k}$ denoting the existing RU mapping values for pre-embedded HP SRs. In addition,  $N^{\text{FRM}},$ and $\Pi^{\text{PRB}}$ denote the number of PRBs in a radio frame and the radio resources occupied by a single PRB, respectively. Overall, the ceiling function $\lceil\cdot\rceil$ restricts L1 RUs only to be assigned to full radio frames and not the fraction of the frame that may be unused by L0.

Thus, using (\ref{eq17}) and (\ref{eq18}), we should enforce the following constraint.
}
\begin{equation}\label{eq19}
    \Pi_{q,\text{TOT}}^{\sigma,\text{UTL}}
    \leq\Pi^{\max}_{q,\sigma}-\Pi_{q,\text{TOT}}^{\sigma,\text{OVH}},\quad\forall \sigma\in\mathcal{S}_q, q\in\mathcal{V}^{\text{P,R}}
\end{equation}
where $\Pi_{q,\sigma}^{\max}$ denote the maximum radio resources at frequency slot $\sigma\in\mathcal{S}_q$ of NR-BS $q\in\mathcal{V}^{\text{P,R}}$.

\subsection{VP Mapping Constraints}
\subsubsection{\textcolor{black}{Mapping Uniqueness}} {\color{black}In this paper, we assume each VP is mapped to at most one PP to maintain spectral contiguity by ensuring a contiguous bandwidth allocation and preventing fragmentation. This approach also reduces overhead for L1 SRs, as bandwidth splitting would require dedicated guardbands and subcarrier for each VP. Thus,
\begin{equation}\label{eq20}
    \sum_{e_{pq}^{(i)}\in\mathcal{E}^{\text{P}}}\min_{l\in\mathcal{L}_{pqi}}\sum_{\lambda\in\mathcal{D}_l}z_{vw}^{l,\lambda} \leq  1,\quad\forall e_{vw}\in \mathcal{E}_{r}, r\in\mathcal{R}
\end{equation}
where the inequality accounts for the fact that no mapping is required for revisited VPs. In addition, $\min(\cdot)$ selects the paths with all the constituent links being mapped. The summation over the set of wavelength slots of link $l\in\mathcal{L}^{\text{P}}$ ensures the spectral contiguity by selecting at most a single slot $\lambda\in\mathcal{D}_l$ on $l$ for VP $e_{vw}\in\mathcal{E}_r$ of SR $r\in\mathcal{R}$.}
\subsubsection{\textcolor{black}{Complete Isolation Mapping}} At L2, a link is only mapped to at most a single SR. Thus,
{\color{black}\begin{equation}\label{eq21}
    f_{l}(\mathcal{E}, \mathcal{R}, \boldsymbol{z}^{l})\leq 1,\quad \forall l\in\mathcal{L}^{\text{P}}
\end{equation}
\begin{equation}\label{eq22}
	\begin{aligned}
		\sum_{\lambda\in\mathcal{D}_l}z_{vw}^{l,\lambda} \leq1-f_l(\mathcal{E}, \mathcal{R}\backslash\{r\}, \boldsymbol{z}^{l}),&\\
        \forall l\in\mathcal{L}^{\text{P}},  e_{vw}\in\mathcal{E}_r,  r\in\mathcal{R}&
	\end{aligned}
\end{equation}
where $\boldsymbol{z}^{l}\triangleq(\sum_{\lambda\in\mathcal{D}_l}z_{vw}^{l,\lambda})_{e_{vw}\in\mathcal{E}_r}^{r\in\mathcal{R}}$.}

\subsubsection{\textcolor{black}{Coordination and Overprovisioning Prevention}} As previously mentioned, we employ tagging for L0 service differentiation, with a fixed-sized header in packets to distinguish slices. Thus, they consume a specific percentage of the allocated bandwidth, denoted by $\psi^{\text{OVH}}_{\text{HDR}}$, which is essential for L0 operation. Given this overhead, we must ensure coordinated bandwidth allocations and mapping decisions. {\color{black}In addition, we should prevent overprovisioning.} Thus,
\begin{equation}\label{eq23}
    \begin{aligned}
	   \epsilon z_{vw}^{l,\lambda}\leq\sum_{e_{pq}^{(i)}\in\mathcal{E}^{\text{P}}_l}b_{vw}^{pqi,l,\lambda}\leq \tfrac{B_{vw}^{\text{REQ}}z_{vw}^{l,\lambda}}{1-\psi^{\text{OVH}}_{\text{HDR}}\mathds{1}_{\left\{\gamma_{r}=0\right\}}},&\\
        \forall l\in\mathcal{L}^{\text{P}},  \lambda\in\mathcal{D}_l,  e_{vw}\in\mathcal{E}_{r},  r\in\mathcal{R}&
    \end{aligned}
\end{equation}
where $\epsilon>0$ is the smallest allocatable bandwidth. A PP's bandwidth is limited by its bottleneck link. Therefore, to efficiently consume bandwidth resources, we ensure that links on a PP receive the same bandwidth, considering their potentially different bandwidth fragmentation as follows.
\begin{equation}\label{eq24}
    \begin{aligned}
	   &\sum_{\lambda\in\mathcal{D}_{l'}}b_{vw}^{pqi,l',\lambda}= \sum_{\lambda\in\mathcal{D}_{l''}}b_{vw}^{pqi,l'',\lambda},\\
        &\forall l',l''\in\mathcal{L}_{pqi},
       l'\neq l'',  e_{pq}^{(i)}\in\mathcal{E}^{\text{P}},  e_{vw}\in\mathcal{E}_{r},  r\in\mathcal{R}
    \end{aligned}
\end{equation}

It is essential that if a VP is mapped to a PP, its end nodes are also mapped. Accordingly,
\begin{equation}\label{eq25}
	\begin{aligned}
	\sum_{e_{pq}^{(i)}\in\mathcal{E}^{\text{P}}}\sum_{\lambda\in\mathcal{D}_{l_1}}b_{vw}^{pqi,l_1,\lambda} \leq \frac{B_{vw}^{\text{REQ}}}{1-\psi^{\text{OVH}}_{\text{HDR}}\mathds{1}_{\left\{\gamma_{r}=0\right\}}}x_v^p,&\\
          \forall e_{vw}\in\mathcal{E}_{r}, p\in\mathcal{V}^{\text{P,S}},  v\in\mathcal{V}^{\text{S}}_{r}, r\in\mathcal{R}&
	\end{aligned}
\end{equation}
\begin{equation}\label{eq26}
	\begin{aligned}
		\sum_{e_{pq}^{(i)}\in\mathcal{E}^{\text{P}}}\sum_{\lambda\in\mathcal{D}_{l_1}}b_{vw}^{pqi,l_1,\lambda}\leq \frac{B_{vw}^{\text{REQ}}}{1-\psi^{\text{OVH}}_{\text{HDR}}\mathds{1}_{\left\{\gamma_{r}=0\right\}}}\sum_{\sigma\in\mathcal{S}_q}\sum_{k=1}^{K_w^{\text{REQ}}}y_w^{q,\sigma,k},&\\
        \forall e_{vw}\in\mathcal{E}_{r}, q\in\mathcal{V}^{\text{P,R}},  w\in\mathcal{V}^{\text{R}}_{r},  r\in\mathcal{R}&
	\end{aligned}
\end{equation}
where $l_1\in\mathcal{L}_{pqi}$ is the link 1 on PP $e_{pq}^{(i)}\in\mathcal{E}^{\text{P}}$. According to (\ref{eq24}), it is unnecessary to repeat (\ref{eq25}) and (\ref{eq26}) for all the links.

{\color{black}When revisiting the end nodes of a VP, there will not be a PP needed to connect those nodes. We first define $h_{vw}^{\text{REV}}$ which is 0 when revisitation occurs for the end nodes of VP $e_{vw}\in\mathcal{E}_r$, i.e., nodes are not mapped to the same PS or NR-BS, and 1 otherwise, as follows.
\begin{displaymath}
    \Tilde{h}_{vw}^{\text{REV}} \triangleq 1-\begin{cases}
    \sum_{p\in\mathcal{V}^{\text{P,S}}}\min(x_v^p,x_w^p) & \text{$v,w\in\mathcal{V}_r^{\text{S}}$} \\
    \sum_{q\in\mathcal{V}^{\text{P,R}}}\min(\sum_{\sigma\in\mathcal{S}_q}\sum_{k=1}^{K_v^{\text{REQ}}}y_v^{q,\sigma,k}\\~~~~~~~~~~~~~~~,\sum_{\sigma\in\mathcal{S}_q}\sum_{k=1}^{K_w^{\text{REQ}}}y_w^{q,\sigma,k}) & \text{$v,w\in\mathcal{V}_r^{\text{R}}$}\\
        0 & \text{otherwise}
\end{cases}
\end{displaymath}
Thus, for VP $e_{vw}\in\mathcal{E}_r$ of SR $r\in\mathcal{R}$, we define the effective difference of the required and allocated bandwidth as
\begin{equation}\label{eq27}
        \hat{h}_{vw}^{\text{B}} \triangleq B_{vw}^{\text{REQ}}h_{vw}^{\text{REV}} - \left(1-\psi^{\text{OVH}}_{\text{HDR}}\mathds{1}_{\left\{\gamma_{r}=0\right\}}\right)\sum_{e_{pq}^{(i)}\in\mathcal{E}^{\text{P}}}\sum_{\lambda\in\mathcal{S}_{l_1}}b_{vw}^{pqi,l,\lambda}
\end{equation}

\noindent To coordinate bandwidth allocation with revisitation,
\begin{equation}\label{eq28}
    \hat{h}_{vw}^{\text{B}} \geq 0,\quad\forall e_{vw}\in\mathcal{E}_r, r\in\mathcal{R}
\end{equation}}
\subsubsection{\textcolor{black}{Maximum Capacity}} For each wavelength slot of each link, the total bandwidth usage must remain within the maximum capacity. {\color{black}Let $B^{\text{OVH}}_{\text{WGB}}$ denote the overhead of each wavelength guard band at L1. Accordingly, the total utilized bandwidth and bandwidth overhead on wavelength slot $\lambda\in\mathcal{D}_l$ of link $l\in\mathcal{L}^{\text{P}}$, denoted by $B_{l,\text{TOT}}^{\lambda,\text{UTL}}$ and $B_{l,\text{TOT}}^{\lambda,\text{OVH}}$, respectively, can be derived as follows.
\begin{equation}\label{eq29}
    \begin{aligned}
    B_{l,\text{TOT}}^{\lambda,\text{UTL}}=&\sum_{r\in\mathcal{R}\cup\mathcal{R}^{\text{HP}}_t}^{\gamma_r\neq 1}\sum_{e_{vw}\in\mathcal{E}_{r}}\sum_{e_{pq}^{(i)}\in\mathcal{E}_l^{\text{P}}}\tilde{b}_{vw}^{pqi,l,\lambda}+&\\
         &B^{\text{WSC}}\sum_{r\in\mathcal{R}\cup\mathcal{R}^{\text{HP}}_t}^{ \gamma_r=1}\sum_{e_{vw}\in\mathcal{E}_{r}}\sum_{e_{pq}^{(i)}\in\mathcal{E}_l^{\text{P}}} \left\lceil\frac{\tilde{b}_{vw}^{pqi,l,\lambda}}{B^{\text{WSC}}}\right\rceil
    \end{aligned}
\end{equation}
\begin{equation}\label{eq30}
    B_{l,\text{TOT}}^{\lambda,\text{OVH}}=\sum_{r\in\mathcal{R}\cup\mathcal{R}^{\text{HP}}_t}^{\gamma_r=1}\sum_{\substack{e_{vw}\in\mathcal{E}_{r}}}\sum_{\substack{e_{pq}^{(i)}\in\mathcal{E}_l^{\text{P}}}}\tilde{z}_{vw}^{l,\lambda}B_{\text{WGB}}^{\text{OVH}}
\end{equation}
where \[
\tilde{b}_{vw}^{pqi,l,\lambda} \triangleq 
\begin{cases} 
b_{vw}^{pqi,l,\lambda}, & \forall e_{vw} \in \mathcal{E}_r, r \in \mathcal{R} \\
B_{vw}^{pqi,l,\lambda}, & \forall e_{vw} \in \mathcal{E}_r, r \in \mathcal{R}^{\text{HP}}_t
\end{cases}
\] \[
\tilde{z}_{vw}^{l,\lambda} \triangleq 
\begin{cases} 
z_{vw}^{l,\lambda}, & \forall e_{vw} \in \mathcal{E}_r, r \in \mathcal{R} \\
Z_{vw}^{l,\lambda}, & \forall e_{vw} \in \mathcal{E}_r, r \in \mathcal{R}^{\text{HP}}_t
\end{cases}
\] with $B_{vw}^{pqi,l,\lambda}$ and $Z_{vw}^{l,\lambda}$ denoting the existing VP bandwidth values and mappings for pre-embedded HP SRs, respectively. In addition, $\mathcal{E}^{\text{P}}_l$, $B^{\text{WSC}}$ and $B^{\text{OVH}}_{\text{WGB}}$ denotes the set of PPs using link $l\in\mathcal{L}^{\text{P}}$, wavelength subcarriers' bandwidth and the overhead of wavelength guard bands, respectively. In (\ref{eq29}), for L1 SRs, the ceiling function $\left\lceil\cdot\right\rceil$ is used to round up the bandwidth to the subcarriers' bandwidth {\color{black}and avoid L0 SRs to get access to the possibly remaining bandwidth of those subcarriers}. For L0, in the first term of (\ref{eq29}), we allocate subcarriers for the cumulative bandwidth of all SRs, as they share subcarriers. We reuse this term for L2 SRs as well.

Therefore, by utilizing (\ref{eq29}) and (\ref{eq30}), we should enforce
\begin{equation}\label{eq31}
    B_{l,\text{TOT}}^{\lambda,\text{UTL}} \leq B^{\max}_{l,\lambda}-B_{l,\text{TOT}}^{\lambda,\text{OVH}},\quad\forall \lambda\in\mathcal{D}_l,  l\in\mathcal{L}^{\text{P}}
\end{equation}
where $B^{\max}_{l,\lambda}$ denote the maximum bandwidth capacity of link $l$ on wavelength slot $\lambda$.}

\subsection{Profit Constraints}
{\color{black}Incorporating isolation levels introduces additional complexity, potentially leading to underutilization of network resources. To mitigate this, in addition to revisitation and migration, one strategy involves permitting controlled QoS violation for LP SRs with a finite-cost penalty in exchange for accepting HP SRs. We need to calculate various costs incurred in the proposed system to derive the profit. These include slice deployment, isolation overheads, migration, and QoS violation costs.

We start by deriving the bandwidth violation cost, denoted by $g_{r}^{\text{B}}$. For SR $r\in\mathcal{R}$, we have}
\begin{equation}\label{eq32}
    g_{r}^{\text{B}}\triangleq\rho_{r}^{\text{B}}\sum_{e_{vw}\in\mathcal{E}_r}\hat{h}_{vw}^{\text{B}}
\end{equation}
where $\rho_r^{\text{B}}\geq0$ is the cost of a unit of bandwidth violation. Regarding each VP $e_{vw}\in\mathcal{E}_r$ of SR $r\in\mathcal{R}$'s RTT delay violation cost, we need to consider {\color{black}that delay is determined by the links and not allocated like bandwidth. Therefore, it can either be greater or smaller than the requested delay. Considering this,} we derive the effective difference between the requested and experienced delay{\color{black}, denoted by $\hat{h}_{vw}^{\text{D}}$,} as
{\color{black}\begin{equation}\label{eq33}
    \hat{h}_{vw}^{\text{D}}\triangleq \left[\sum_{e_{pq}^{(i)}\in\mathcal{E}^{\text{P}}}\Big(D_{pqi}^{\text{RTT}}\min_{l\in\mathcal{L}_{pqi}}\sum_{\lambda\in\mathcal{D}_l}z_{vw}^{l,\lambda}\Big)-h_{vw}^{\text{D}}\right]^+
\end{equation}}
where $D_{pqi}^{\text{RTT}}$ is the RTT delay of PP $e_{pq}^{(i)}\in\mathcal{E}^{\text{P}}$ and $h_{vw}^{\text{D}} \triangleq D_{vw}^{\text{REQ}} + M(1- \Tilde{h}_{vw}^{\text{REV}})$ with $M$ as a sufficiently large number to cover the case of revisitation. $[\cdot]^+\triangleq \max(0, \cdot)$ ensures that only positive delay violations are considered. Thus, we can derive the delay violation cost of SR $r\in\mathcal{R}$, {\color{black} denoted by $g_{r}^{\text{D}}$,} as
\begin{equation}\label{eq34}
        g_{r}^{\text{D}}\triangleq
	\rho_{r}^{\text{D}}\sum_{e_{vw}\in\mathcal{E}_{r}}\hat{h}_{vw}^{\text{D}}
\end{equation}

Let $\rho_r^{\text{C}}$ and $\rho_r^{\text{K}}$ denote the cost of a unit of MIPS and radio resource violation, respectively. We can then derive the MIPS and radio violation costs of $r\in\mathcal{R}${\color{black}, denoted by $g_{r}^{\text{C}}$ and $g_{r}^{\text{K}}$}, respectively, as
\begin{equation}\label{eq35}
        g_{r}^{\text{C}}\triangleq
	\rho_{r}^{\text{C}}\sum_{v\in\mathcal{V}_r^{\text{S}}}\left(C_v^{\text{REQ}}-\sum_{p\in\mathcal{V}^{\text{P,S}}}c_v^p\right)
\end{equation}
\begin{equation}\label{eq36}
        g_{r}^{\text{K}}\triangleq
	\rho_{r}^{\text{K}}\sum_{w\in\mathcal{V}_r^{\text{R}}}\left(K_w^{\text{REQ}}-\sum_{q\in\mathcal{V}^{\text{P,R}}}\sum_{\sigma\in\mathcal{S}_q}\sum_{k=1}^{K_w^{\text{REQ}}}ky_w^{q,\sigma,k}\right)
\end{equation}

Now, by adding (\ref{eq32}) and (\ref{eq34})--(\ref{eq36}), we can derive SR $r\in\mathcal{R}$'s total QoS violation cost{\color{black}, denoted by $g_{r}^{\text{VIO}}$,} as 
\begin{equation}\label{eq37}
    \begin{aligned}
	g_{r}^{\text{VIO}}\triangleq g_{r}^{\text{C}}+g_{r}^{\text{K}}+g_{r}^{\text{B}}+g_{r}^{\text{D}}
 \end{aligned}
\end{equation}

The next cost is on deployment and the associated overheads. First, let $\Omega$, $\Phi$, and $\Theta$ denote the cost of a unit of MIPS, PRB, and bandwidth. Thus, the deployment cost of {\color{black}L0 and L1} SR $r\in\mathcal{R}${\color{black}, denoted by $g^{\text{DEP,L0}}_{r}$ and $g^{\text{DEP,L1}}_{r}$, respectively,} is
{\color{black}\begin{displaymath}
    \begin{aligned}
            g^{\text{DEP, L0}}_{r}\triangleq&\Phi\sum_{\substack{w\in\mathcal{V}^{\text{R}}_{r}}}^{\gamma_r= 0}\sum_{q\in\mathcal{V}^{\text{P,R}}}\sum_{\sigma\in\mathcal{S}_q}\sum_{k=1}^{K_w^{\text{REQ}}} ky_{w}^{q,\sigma,k} +\Omega\sum_{v\in\mathcal{V}_{r}^{\text{S}}}^{\gamma_r=0}\sum_{p\in\mathcal{V}^{\text{P,S}}} c_{v}^{p}\\
            &+\Theta\sum_{e_{vw}\in\mathcal{E}_{r}}^{\gamma_r=0}\sum_{e_{pq}^{(i)}\in\mathcal{E}^{\text{P}}}\sum_{l\in\mathcal{L}_{pqi}}\sum_{\lambda\in\mathcal{D}_l} b_{vw}^{pqi,l,\lambda}
    \end{aligned}
\end{displaymath}
\begin{displaymath}
    \begin{aligned}
            &g^{\text{DEP, L1}}_{r}\triangleq\Phi N^{\text{FRM}}\sum_{\substack{w\in\mathcal{V}^{\text{R}}_{r}}}^{\gamma_r= 1}\sum_{q\in\mathcal{V}^{\text{P,R}}}\sum_{\sigma\in\mathcal{S}_q}\sum_{k=1}^{K_w^{\text{REQ}}} \left\lceil\frac{ky_{w}^{q,\sigma,k}}{N^{\text{FRM}}}\right\rceil + \\
            &\Omega\sum_{v\in\mathcal{V}_{r}^{\text{S}}}^{\gamma_r=1}\sum_{p\in\mathcal{V}^{\text{P,S}}} c_{v}^{p} + \Theta B^{\text{WSC}}\sum_{e_{vw}\in\mathcal{E}_{r}}^{\gamma_r=1}\sum_{e_{pq}^{(i)}\in\mathcal{E}^{\text{P}}}\sum_{l\in\mathcal{L}_{pqi}}\sum_{\lambda\in\mathcal{D}_l} \left\lceil\frac{b_{vw}^{pqi,l,\lambda}}{B^{\text{WSC}}}\right\rceil
    \end{aligned}
\end{displaymath}

For L2 SRs, the deployment cost should consider the allocation of all resources in the mapped PS, NR-BS, and PP. In addition, this should account for when multiple virtual nodes of the same L2 SR are using the same physical entity and avoid overstating the deployment cost. Therefore, the deployment cost, denoted by $g^{\text{DEP,L2}}_{r}$, can be derived as follows.
\color{black}\begin{displaymath}
    \begin{aligned}
            &g^{\text{DEP, L2}}_{r}\triangleq\Phi\sum_{q\in\mathcal{V}^{\text{P,R}}}\frac{\Pi^{\max}_{q,0}}{\Pi^{\text{PRB}}}\max_{\substack{w\in\mathcal{V}^{\text{R}}_{r} \\ \gamma_r = 2}}\sum_{k=1}^{K_w^{\text{REQ}}} y_w^{q,0,k}+\\
            &\Omega\sum_{p\in\mathcal{V}^{\text{P,S}}} C_p^{\max}\max_{\substack{v\in\mathcal{V}_{r}^{\text{S}} \\ \gamma_r = 2}}x_{v}^{p} + \Theta\sum_{l\in\mathcal{L}}B^{\max}_{l,0} \max_{\substack{e_{vw}\in\mathcal{E}_{r} \\ \gamma_r = 2}}z_{vw}^{l,0}
    \end{aligned}
\end{displaymath}

According to Fig. \ref{isolation}, overheads are not necessarily specific to individual SRs, e.g., hypervisor or host OS. Therefore, we derive the total overhead costs{\color{black}, denoted by $g^{\text{OVH}}_{\text{TOT}}$,} as follows. 
\begin{displaymath}
    \begin{aligned}
            g^{\text{OVH}}_{\text{TOT}}\triangleq\Phi\sum_{q\in\mathcal{V}^{\text{P,R}}}\sum_{\sigma\in\mathcal{S}_q}\Pi_{q,\text{TOT}}^{\sigma,\text{OVH}} + \Omega\sum_{p\in\mathcal{V}^{\text{P,S}}}C_{p,\text{TOT}}^{\text{OVH}}+&\\
            \Theta\sum_{l\in\mathcal{L}^{P}}\sum_{\lambda\in\mathcal{D}_l}B_{l,\text{TOT}}^{\lambda,\text{OVH}}&
    \end{aligned}
\end{displaymath}}

Migration of RU and NF nodes of pre-embedded slices must be associated with a cost, {\color{black} denoted by $g^{\text{MIG}}_{\hat{r}}$ for SR $\hat{r}\in\mathcal{R}_t$,} that depends on the slice priority. Thus,
\begin{displaymath}
    \begin{aligned}
		&g^{\text{MIG}}_{\hat{r}}\triangleq\beta_{\hat{r}}\sum_{v\in\mathcal{V}_{\hat{r}}^{\text{S}}}\sum_{p\in\mathcal{V}^{\text{P,S}}}x_v^p(1-X_v^{p}) + \\
        &\beta_{\hat{r}}\sum_{w\in\mathcal{V}_{\hat{r}}^{\text{R}}}\sum_{q\in\mathcal{V}^{\text{P,R}}}\Bigg(\sum_{\sigma\in\mathcal{S}_q}\sum_{k=1}^{K_w^{\text{REQ}}}y_w^{q,\sigma,k}\Bigg)\Bigg(1-\sum_{\sigma\in\mathcal{S}_q}\sum_{k=1}^{K_w^{\text{REQ}}}Y_w^{q,\sigma,k}\Bigg)
	\end{aligned}
\end{displaymath}
where $Y_{w}^{q,\sigma,k}$ denote the existing mapping decisions of RU $w\in\mathcal{V}^{\text{R}}_{\hat{r}}$ of SR ${\hat{r}}\in\mathcal{R}_t$, respectively.

The InP wants to ensure that the added profit after admitting $r_t$ at time step $t$ is higher than a minimum $P^{\min}$. {\color{black}With $g^{\text{DEP}}_{r}\triangleq g^{\text{DEP, L0}}_{r}+g^{\text{DEP, L1}}_{r}+g^{\text{DEP, L2}}_{r}$, we first define the InP's obtained profit at time step $t$ as follows.}
{\color{black}\begin{equation*}
    \begin{aligned}
        P_t\triangleq\sum_{r\in\mathcal{R}\cup\mathcal{R}^{\text{HP}}_t}R_{r} - \sum_{r\in\mathcal{R}}\left(g^{\text{DEP}}_{r} - g^{\text{VIO}}_{r}\right) -\sum_{\hat{r}\in\mathcal{R}_t}g^{\text{MIG}}_{\hat{r}}- g^{\text{OVH}}_{\text{TOT}}&\\
        - g^{\text{DEP}}_{\text{HP},t}&
    \end{aligned}
\end{equation*}}
{\color{black}where $g^{\text{DEP}}_{\text{HP},t}$ is the total deployment cost of pre-embedded HP SRs at time step $t$.}
{\color{black}\begin{equation}\label{eq38}
    P_t - P_{t-1}\geq P^{\min}
\end{equation}}
where $P_{t-1}$ is given from time step $t-1$.

\subsection{5G Isolation-Aware Network Slicing (5G-INS) Problem}
We consider the objective to be the cumulative profit the InP expects to generate if it admits $r_t$. Therefore, we can define the 5G-INS problem as follows.
\begin{subequations}
	\begin{alignat}{4}
		\displaystyle{\maximize_{\boldsymbol{x}, \boldsymbol{y}, \boldsymbol{z}, \boldsymbol{c}, \boldsymbol{b}}}\quad &  P_t\\
		\textrm{subject to} \quad & (\ref{eq1}), (\ref{eq2}), (\ref{eq4})\text{--}(\ref{eq9}), (\ref{eq15}), (\ref{eq16}), (\ref{eq19})\text{--}(\ref{eq26}),\\
        &(\ref{eq28}), (\ref{eq31}), (\ref{eq38})  \label{p1b}
	\end{alignat}
\end{subequations}

\subsection{Analysis and Simplification}
5G-INS is a mixed-integer programming (MIP) problem that contains non-linear, discontinuous functions such as the indicator function. {\color{black}This alone does not mean that 5G-INS is intractable. However, }the following theorem applies.
\begin{theorem}\label{thm1}
    5G-INS is an NP-hard problem.
\end{theorem}
\begin{proof}
Refer to Appendix \ref{appa}.
\end{proof}
Therefore, 5G-INS is intractable, and we cannot trivially tackle 5G-INS due to NP-hardness. However, we present the Theorem below{\color{black}, which makes the reduction of 5G-INS to a tractable form possible.}
\begin{theorem}\label{thm2}
    We reduce 5G-INS into a linear program by relaxing the binary input variables to the range $[0,1]$.
\end{theorem}
\begin{proof}
Refer to Appendix \ref{appb}.
\end{proof}
\begin{figure}
	\centering
	\includegraphics[width=0.85\linewidth]{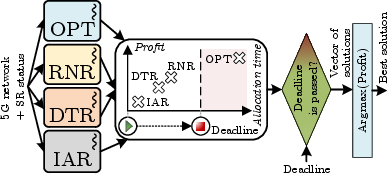}
	\caption{\textcolor{black}{Flow diagram of the 5Guard framework for a 5G-INS instance. The ensemble algorithms, e.g., IAR, DTR, RNR, and OPT, operate concurrently on separate threads, processing shared input data. If the resource allocation deadline is reached, all algorithm executions are halted, and the best feasible solution is selected.
    }}\label{5guard}
\end{figure}
{\color{black}Another aspect of 5G-INS to analyze is its integrality gap, i.e., the maximum ratio of the original problem's optimal solution to the relaxed problem's, which can reveal how rounding the binary variables affects the optimal solution. We present the following Theorem.}
\begin{theorem}\label{thm3}
    The integrality gap of 5G-INS is unbounded.
\end{theorem}
\begin{proof}
Refer to Appendix \ref{appc}.
\end{proof}

In the following section, we will leverage Theorem \ref{thm2} and \ref{thm3} to propose a framework that effectively addresses the complexity of 5G-INS.

\section{5Guard Framework}
{\color{black}In this section, we propose 5Guard, a framework to address 5G-INS. We then introduce four algorithms with distinct characteristics for use within 5Guard and analyze the complexity.}
\subsection{\textcolor{black}{Proposed Framework and Algorithms}}
{\color{black}According to Theorem \ref{thm2},} as 5G-INS can be reduced to an LP for relaxed binary variables, a range of algorithms can be chosen. While state-of-the-art algorithms sometimes offer guarantees for worst-case scenarios, this does not ensure superior performance in all situations. Specifically, algorithms with better worst-case guarantees may not always outperform others in average cases, where performance can vary based on the context. Moreover, modern integer LP (ILP) solvers, such as Gurobi, leverage advanced techniques like branch-and-cut combined with the barrier method \cite{gurobi} to exploit multiple cores, enhancing computational efficiency while guaranteeing optimality. This implies that optimal solutions may be achievable for slices with lenient resource allocation deadlines. Thus, individual algorithms have tradeoffs in execution time or optimality.

\begin{algorithm}[t]
	\DontPrintSemicolon
	\caption{5Guard Framework}\label{alg1}
        \small
	\SetKwProg{Coblock}{cobegin}{ }{coend}
 \KwIn{$\mathcal{V}^{\text{P,S}},\mathcal{V}^{\text{P,R}}, \mathcal{E}^{\text{P}}, \mathcal{L}^{\text{P}}, \mathcal{R}_t, r_t, P^{\min}, \Pi^{\text{PRB}}, N^{\text{FRM}}, t$}
	$\mathcal{A}\triangleq\{A_i | A_i\text{ is an optimization algorithm}\}$;\;
    $T^{\text{STA}}=\text{time}()$\;
    Instantiate $|\mathcal{A}|$ number of processing threads\;
    \Coblock{$A_i$}{
        Execute $A_i$ on the input to solve 5G-INS at time step $t$\;
    }
    \While{All $A_i$s are not finished AND $\text{time}()-T^{\text{STA}}<T_{r_t}$}{continue\;}

    $\mathcal{A}^{\text{RET}}\triangleq\{A_i|A_i \text{ returned a feasible solution}\}$\;

    \If{$\mathcal{A}^{\text{RET}}=\varnothing$}{
        \Return{Infeasible}
    }
    $i^\ast = \text{argmax}(\text{profit}(\mathcal{A}^{\text{RET}}))$\;
    $\boldsymbol{x}^\ast,\boldsymbol{y}^\ast,\boldsymbol{z}^\ast,\boldsymbol{c}^\ast,\boldsymbol{b}^* = \text{argmax}(A^{\text{RET}}_{i^\ast})$\;
	\KwOut{$\boldsymbol{x}^*$, $\boldsymbol{y}^*$, $\boldsymbol{z}^*$, $\boldsymbol{c}^*$, $\boldsymbol{b}^*$}
\end{algorithm}

{\color{black}As illustrated in Fig. \ref{5guard}, }to address this issue, we propose 5Guard, a framework that leverages an ensemble of optimization algorithms and exploits parallel processing to identify the best solution within the resource allocation deadline. The pseudo-code of 5Guard for each time step $t$ is provided in Algorithm \ref{alg1}. Here, time steps represent the iterations of 5Guard, which is distinct from execution time. In Line 1, we define the set $\mathcal{A}$ as the ensemble of optimization algorithms. We first measure the start time in Line 2 with the function $\text{time}()$. In Line 3, we instantiate $|\mathcal{A}|$ processing threads to run the algorithms in parallel. In Lines 4 and 5, ``cobegin $A_i$'' indicates the parallel block of code parameterized with algorithm $A_i$ \cite{iotj232,wcl24}. In Lines 6 and 7, 5Guard waits until all the algorithms complete execution or the elapsed time exceeds the resource allocation deadline $T_{r_t}$ of SR $r_t$. In Line 8, we store the algorithms that return a feasible solution in set $\mathcal{A}^{\text{RET}}$. If $\mathcal{A}^{\text{RET}}$ is empty, it implies that no algorithm could reach a feasible solution or finish in time. Thus, 5Guard returns infeasible, hence rejecting $r_t$. Otherwise, it chooses the index $i^\ast$ that has achieved the maximum profit. It finally returns $A_{i^\ast}^{\text{RET}}$'s decisions.

For this work{\color{black}, as a proof-of-concept,} we design set $\mathcal{A}$ from four custom algorithms: 1) Isolation-aware ranking (IAR), 2) deterministic rounding (DTR), 3) randomized rounding (RNR), and 4) exact (OPT), where OPT uses Gurobi's solver based on branch-and-cut and barrier method. These algorithms present a wide range of complexity and approximation ratios. Based on the unbounded integrality gap of 5G-INS {\color{black}proven in Theorem \ref{thm3}}, only relying on rounding techniques can be limiting.

\begin{algorithm}[t]
	\DontPrintSemicolon
	\caption{IAR Algorithm}\label{alg2}
        \small
	\SetKwProg{Coblock}{cobegin}{ }{coend}
 \KwIn{$\mathcal{V}^{\text{P,S}},\mathcal{V}^{\text{P,R}}, \mathcal{E}^{\text{P}}, \mathcal{L}^{\text{P}}, \mathcal{R}_t, r_t, P^{\min}, \Pi^{\text{PRB}}, N^{\text{FRM}}, t$}
        $\boldsymbol{x}^{\ast}=\boldsymbol{y}^{\ast}=\boldsymbol{z}^{\ast}=\boldsymbol{0}$\;
        Maintain the mappings for pre-embedded HP SRs from $t-1$\;
        $\mathcal{R}^{\text{M}}= \mathcal{R}$\;
        \While{$\mathcal{R}^{\text{M}}\neq \emptyset$}{
            $r^\ast = \text{argmax}_{r\in\mathcal{R}^{\text{M}}}R_r(\gamma_r+1)(\beta_r+1)$\;
            $\mathcal{V}_{r^\ast}^{\text{R,M}}= \mathcal{V}_{r^\ast}^{\text{R}}$\;
            $\mathcal{V}^{\text{P,R,M}}=\mathcal{V}^{\text{P,R}}\backslash \{\text{Completely isolated NR-BSs}\}$\;
            \While{$\mathcal{V}_{r^\ast}^{\text{R,M}}\neq \emptyset$}{
                \lIf{$\mathcal{V}^{\text{P,R,M}}=\emptyset$}{
                    \Return infeasible
                }
                Set $w^\ast$, $q^\ast$, $f^\ast$, and $k^\ast$ based on (\ref{eq42})\;
                \lIf{$k^\ast = 0$}{
                    \Return infeasible
                }
                $(y_{w^\ast}^{q^\ast,f^\ast,k^\ast})^\ast = 1$\;
                \lIf{$\gamma_{r^\ast} = 2$}{
                    Mark $q^{\ast}$ as completely isolated
                }
                
                $\mathcal{V}_{r^\ast}^{\text{R,M}}=\mathcal{V}_{r^\ast}^{\text{R,M}}\backslash \{w^\ast\}$
            }
            Repeat Lines 6-12 for PSs to obtain $\boldsymbol{x}^{\ast}$\;
            \While{$\mathcal{E}_{r^\ast}^{\text{M}}\neq \emptyset$}{
                $e_{vw}^{\ast} = \text{argmax}_{e_{vw}\in\mathcal{E}_{r^\ast}}B_{vw}^{\text{REQ}}$\;
                $\mathcal{E}^{\text{P,M}}={\text{set of PPs connecting mapped $v$ and $w$}}$\;
                \If{$\Tilde{h}_{vw}^{\text{REV}}=1$}{
                    $e_{pq}^{(i)^\ast} = \text{argmax}_{e_{pq}^{(i)}\in\mathcal{E}^{\text{P,M}}}(\min_{l\in\mathcal{L}_{pqi},\lambda\in\mathcal{D}_l}{B_{l,\lambda}^{\text{REM}}})$\;
                    \For{$l^\ast\in\mathcal{L}_{(pqi)^\ast}$}{
                        $\lambda^\ast = \text{argmax}_{\lambda\in\mathcal{D}_{l^\ast}}(B_{{l^\ast},\lambda}^{\text{AVA}})$\;
                        $\left(z_{(vw)^\ast}^{l^\ast,s^\ast}\right)^\ast=1$\;
                    }
                }
                
                $\mathcal{E}_{r^\ast}^{\text{M}}=\mathcal{E}_{r^\ast}^{\text{M}}\backslash \{e_{vw}^{\ast}\}$\;
            }
            $\mathcal{R}^{\text{M}}=\mathcal{R}^{\text{M}}\backslash\{r^\ast\}$\;
        }
        Solve 5G-INS for $(\boldsymbol{x},\boldsymbol{y},\boldsymbol{z})=(\boldsymbol{x}^\ast,\boldsymbol{y}^\ast,\boldsymbol{z}^\ast)$ and obtain $\boldsymbol{c}^\ast,\boldsymbol{b}^*$\;
	\KwOut{$\boldsymbol{x}^*$, $\boldsymbol{y}^*$, $\boldsymbol{z}^*$, $\boldsymbol{c}^*$, $\boldsymbol{b}^*$}
\end{algorithm}

The IAR algorithm is outlined in Algorithm \ref{alg2}. We first define $\boldsymbol{x}^\ast$, $\boldsymbol{y}^\ast$, and $\boldsymbol{z}^\ast$ as placeholder vectors for optimal values of $\boldsymbol{x}$, $\boldsymbol{y}$, and $\boldsymbol{z}$. Next, we keep the mappings of the pre-embedded HP SRs in Line 2 based on time step $t-1$. In Line 5, we iteratively rank the remaining SRs, defined as $\mathcal{R}^{\text{M}}$, by profit, isolation level, and priority. {\color{black}In Line 6 and 7, considering the complete isolation required by L2, we keep track of the remaining RUs of  {\color{black}selected SR $r^\ast$} and NR-BSs, defined as $\mathcal{V}_{r^\ast}^{\text{R,M}}$ and $\mathcal{V}^{\text{P,R,M}}$, respectively. In Line 10, to avoid bottlenecks, we select RUs and non-isolated NR-BSs with the highest radio resource demand and remaining radio resources, respectively. In addition, the selected RU receives the maximum possible PRBs. Accordingly, the selected RU, NR-BS, frequency slot, and number of PRBs, denoted by $w^\ast$, $q^\ast$, $f^\ast$, $k^\ast$, respectively, will be obtained as follows.
\begin{equation}\label{eq42}
    \begin{aligned}
        w^\ast &= \text{argmax}_{w\in\mathcal{V}_{r^\ast}^{\text{R,M}}}(K_w^{\text{REQ}})\\
        q^\ast &= \text{argmax}_{q\in\mathcal{V}^{\text{P,R,M}}}(\max_{f\in\mathcal{F}_q}\Pi_{q,f}^{\text{REM}})\\
        f^\ast &= \text{argmax}_{f\in\mathcal{F}_{q^\ast}}(\Pi_{q^\ast,f}^{\text{REM}})\\
        k^\ast &= \min(K_{w^\ast}^{\text{REQ}},\Pi_{q^\ast,f^\ast}^{\text{REM}})
    \end{aligned}
\end{equation}
where 
\begin{equation*}
    \Pi_{q,f}^{\text{REM}}\triangleq\Pi_{q,f}^{\max}-\sum_{r\in\mathcal{R}}\sum_{w\in\mathcal{V}^{\text{R}}_r}\sum_{q\in\mathcal{V}^{\text{P,R}}}\sum_{f\in\mathcal{F}_q}(y_{w}^{q,f,k})^\ast
\end{equation*}
denotes the remaining radio resource on frequency slot $f$ of NR-BS $q$ at each iteration.}

If L2 isolation is required, {\color{black}we mark the selected NR-BS from the remaining set $V^{\text{P,R,M}}$ in Line 13 to be excluded from the next iteration in Line 7.} We similarly repeat these steps for PSs {\color{black}in Line 15}, assuming a fully connected 5G network. {\color{black}We select VPs based on bandwidth demand, similar to RUs and NFs in Line 17. However, we first have to check whether revisitation has occurred for the end nodes of the selected VP.} The revisitation of end nodes results in skipping the VP. Otherwise, we choose greedily the PP, links, and wavelength slots as outlined in Lines 20--23. In case of infeasibility, the algorithm rejects $r_t$ as pre-admitted slices cannot be dismissed. Otherwise, IAR solves the 5G-INS problem with the obtained mappings and outputs the solution.

{\color{black}Finally, for rounding algorithms, i.e., DTR and RNR, we first use Theorem \ref{thm2} to solve 5G-INS with relaxed binary variables in polynomial time. For deterministic rounding, we utilize a fixed limit to round the potentially fractional values for binary mapping variables, keeping in mind that rounded values should not violate any constraint. For randomized rounding, we treat the solution of the relaxed problem for binary variables as probabilities for rounding, which again should adhere to the constraints.}

\subsection{\textcolor{black}{Complexity Analysis}}
{\color{black}Since 5Guard is a parallel algorithm, we should analyze both \textit{depth} and \textit{work} for its complexity \cite{alg82}. \textit{Depth} is defined as the time complexity of the longest series of computations executed sequentially. \textit{Work} denotes the total number of primitive works executed by the parallel processor.} As the allocation time is limited to $T_{r_t}$ for SR $r_t$, {\color{black}depth} is of the order $\mathcal{O}(T_{r_t})$. If $T_{r_t}$ is unrestricted, the {\color{black}depth} is limited by the slowest performing algorithm. If $\mathcal{A}$ includes an exact algorithm such as branch-and-cut, 5Guard will continue executing until the optimal solution is found. {\color{black}Let $T^\max$ denote the highest time complexity among the algorithms in set $\mathcal{A}$.  In the worst case, all the algorithms in $\mathcal{A}$ finish their execution. Therefore, the depth complexity will be of $\mathcal{O}(T^\max)$, and work complexity will be of $\mathcal{O}(\sum_{A_i\in\mathcal{A}}T^\max)=\mathcal{O}(T^\max)$. This means that having lower complexity algorithms to support challenging requirements does not significantly impact the overall processing work done by the decision-making processor. Although this can be of exponential order depending on the algorithms in $\mathcal{A}$, we must note that the depth and work complexity are fully controllable by $T_{r_t}$ to achieve the best solution in time.}

The first three algorithms discussed rely on solving LP forms of the 5G-INS problem. This paper uses barrier methods, a subset of interior-point methods characterized by a logarithmic barrier function, to solve the problem. This method exhibits a computational time complexity of $\mathcal{O}(M^{1.5}.N^{2})$ \cite{siam}, where $M$ and $N$ denote the number of constraints and variables, respectively. This complexity also depends on parameters controlling input and solution precision. According to constraint (\ref{eq23}) and considering {\color{black}that the 5G substrate network is often a complete graph, i.e., number of edges larger than the nodes,} $M$ and $N$ will both be in the order of $\mathcal{O}(L.S.E.R)$, where $L\triangleq|\mathcal{L}^{\text{P}}|$, $S\triangleq\max_{l\in\mathcal{L}^{\text{P}}}{\mathcal{D}_l}$, $E\triangleq\max_{r\in\mathcal{R}}{\mathcal{E}_r}$, and $R\triangleq|\mathcal{R}|$. {\color{black}This means that these three algorithms are of polynomial-time complexity, making them scalable with the size of the substrate network and SR load. Note that the reductions made in Theorem \ref{thm2} and the max operations in (\ref{eq42}) do not change the time complexity.} Although the order of complexity is the same for these algorithms, the isolation-aware ranking algorithm proves to be the fastest in general, due to having {\color{black}considerably} fewer constraints and variables (i.e., $\boldsymbol{b}$ and $\boldsymbol{c}$) in the relaxed problem. {\color{black}This makes it suitable for SRs requiring a fast resource allocation time, such as MC services.}
\begin{table}
        \scriptsize
	\centering
	\caption{Key parameters setting}
	\label{table2}
	\begin{tabular}{p{0.63cm} | p{0.7cm} || p{1.6cm} | p{1.7cm} || p{0.6cm} | p{0.63cm}}
		\hline
		{\footnotesize \textbf{Par.}} & {\footnotesize \textbf{Value}} & {\footnotesize \textbf{Par.}} & {\footnotesize \textbf{Value}} & {\footnotesize \textbf{Par.}} & {\footnotesize \textbf{Value}} \\
		\hline
        \hline
		$\Pi^{\text{PRB}}$ & 180 & $\Pi^{\text{OVH}}_{\text{FGB}}$ & 693 & $N^{\text{FRM}}$ & 10\\
        \hline
        $\epsilon$ & 1Hz & $\Pi_{q,0}^{\max}$ & 500,040 & $\psi_{\text{HDR}}^{\text{OVH}}$ & 0.005\\
        \hline
        $M$ & 1sec. & $B^{\text{WSC}}$, $B^{\text{WGB}}$ & 12.5 GHz & $B_{l,0}^{\max}$ & 4.8 THz\\
        \hline
        $C_{\text{OVH}}^{\text{GOS}}$, & 25,000 & $C_p^{\max}$ & 834,797~MIPS,& $\delta$ & $10^{-6}$\\
        $C_{\text{OVH}}^{\text{K8S}}$ & MIPS & (MEC, Cloud) & 8,347,970~MIPS & & MIPS\\
        \hline
        $\chi_{\text{S}}^{\max}$ & 6 & $C_{\text{OVH}}^{\text{HHO}}$ & 100,000 MIPS &\\
        \hline
	\end{tabular}
\end{table}
\section{Evaluation}
{\color{black}In this section, we evaluate the performance of 5Guard\footnote{5Guard's repository is available at \href{https://github.com/mehdibolourian/5Guard}{github.com/mehdibolourian/5Guard}, allowing for result reproduction and experiment extensions.} by analyzing various metrics across different isolation levels. We also examine the overheads of integrating isolation into 5G NS across different resource domains. Finally, we discuss the key remarks and limitations of this study.}
\subsection{Simulation Setup}

For the 5G system, we follow the values used in the specifications \cite{3gpp242, rfc3031, itu18}, as listed in Table \ref{table2}. These values are derived based on the following assumptions. PRBs have 12 subcarriers, 15 KHz spacing, and 1 ms interval \cite{3gpp242}. Radio frame intervals are set to 10 ms \cite{3gpp242}, and NR-BS bandwidth to 50 MHz \cite{3gpp242}. MEC servers use AMD EPYC 9684X, and 10 such servers are deployed in the cloud. Based on their benchmark results, each server outputs 834,797 MIPS \cite{cpu25}. The wavelength range is 191.3 THz-196.1 THz as used in the industry \cite{itu18}. The MPLS protocol employs 32-bit headers \cite{rfc3031} and a maximum IPv4 packet size of 65535 bits. {\color{black}Our resource allocator testbed utilizes a server-grade Intel Xeon Silver 4314 CPU with 64 threads and 1 TB of RAM to simulate a 5G InP setup.}

\begin{figure}
    \centering

    \subfloat[Synthetic]{%
        \includegraphics[width=0.22\textwidth]{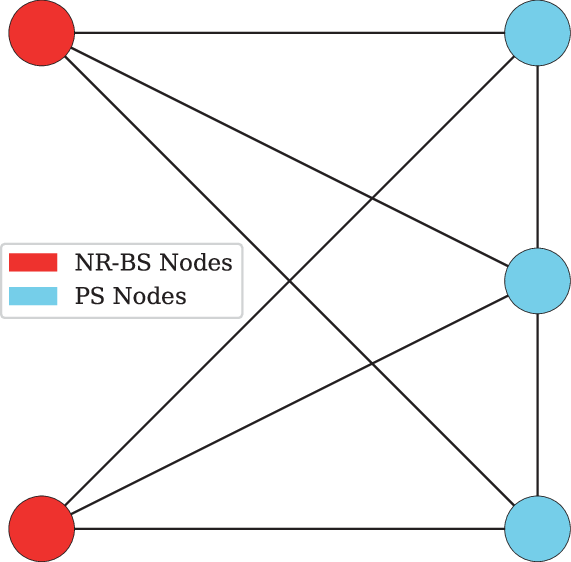}
        \label{figa-setup}
    }
    \hfill
    \subfloat[BRAIN's 5G]{%
        \includegraphics[width=0.22\textwidth]{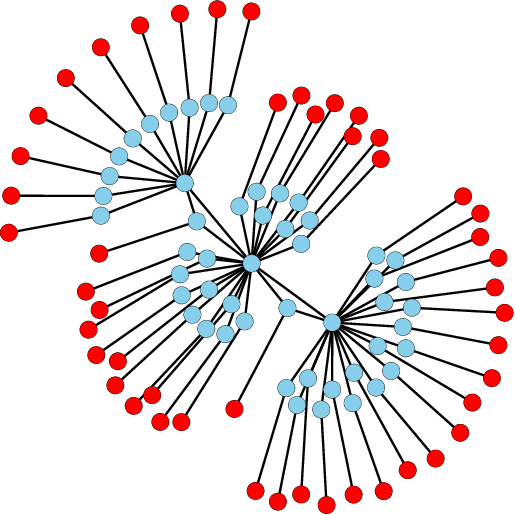}
        \label{figb-setup}
    }

    \caption{5G infrastructure topologies used in the evaluation}
    \label{fig-setup}
\end{figure}

In our simulation, we evaluate 5Guard’s performance and analyze the impact of isolation on 5G NS in resource-critical scenarios and large-scale topologies. For the first goal, as shown in Fig. \ref{figa-setup}, we design a synthetic topology resembling a small MEC/telco cloud network, fully connected to NR-BSs, with a small amount of resources. For large-scale evaluation, as shown in Fig. \ref{figb-setup}, we use a scaled version of the Berlin Research Area Information Network (BRAIN) \cite{brain} as a real-world network topology for our 5G MEC/telco cloud network, featuring 51 PSs, 48 NR-BSs, and 152 PPs. {\color{black}The BRAIN's 5G topology follows the resource values in Table \ref{table2}, while the synthetic topology assumes only 20\% of these resources to model a more resource-critical scenario.}

\begin{figure}
	\centering
	\includegraphics[width=0.9\linewidth]{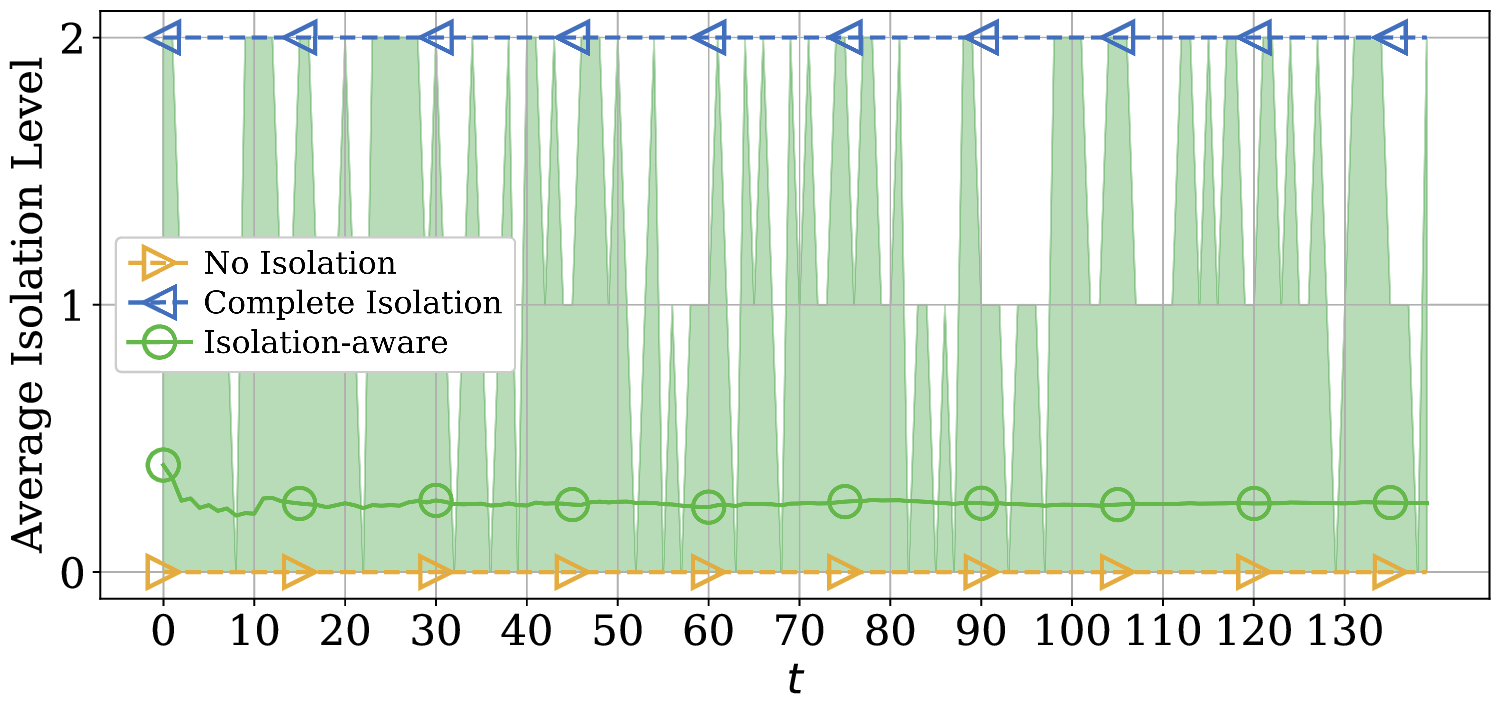}
	\caption{\textcolor{black}{The average SR's isolation level fed to isolation-aware approaches, the approach with no isolation, and complete isolation at each time step, with the shaded area showing the range of absolute isolation levels across the iterations
    }}\label{fig-srisol}
\end{figure}

\begin{figure*}
    \centering
    \subfloat[]{%
        \includegraphics[width=0.24\textwidth]{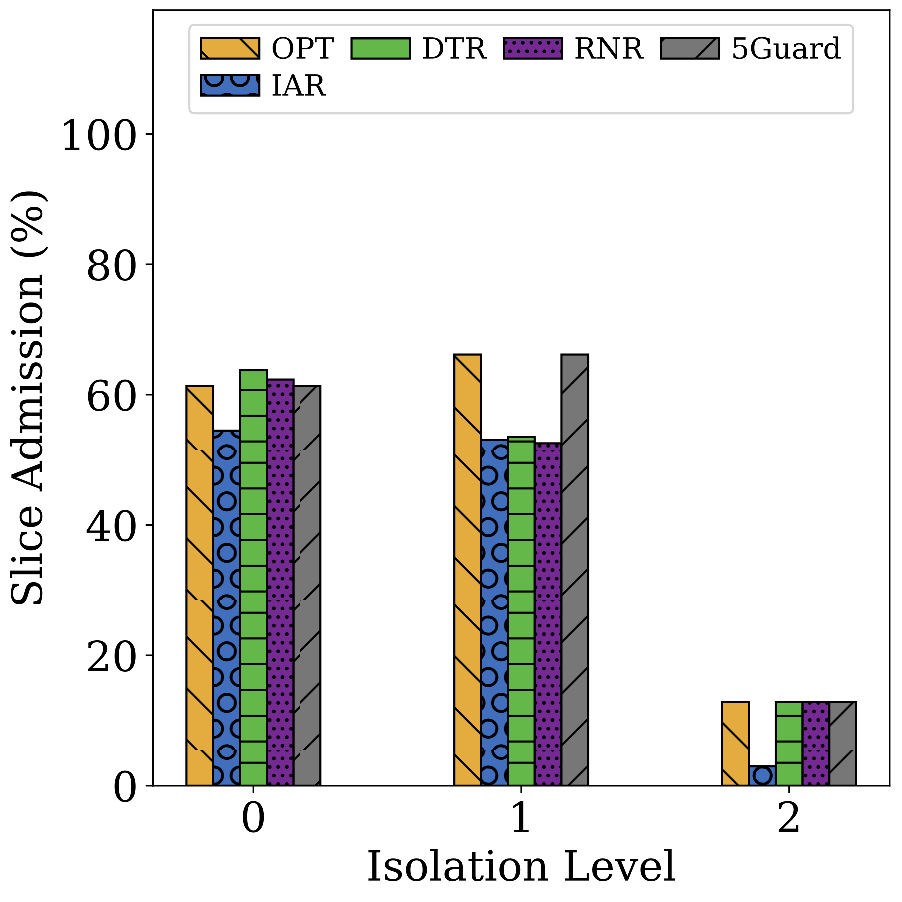}
        \label{figa-sim1}
    }
    \hfill
    \subfloat[]{%
        \includegraphics[width=0.24\textwidth]{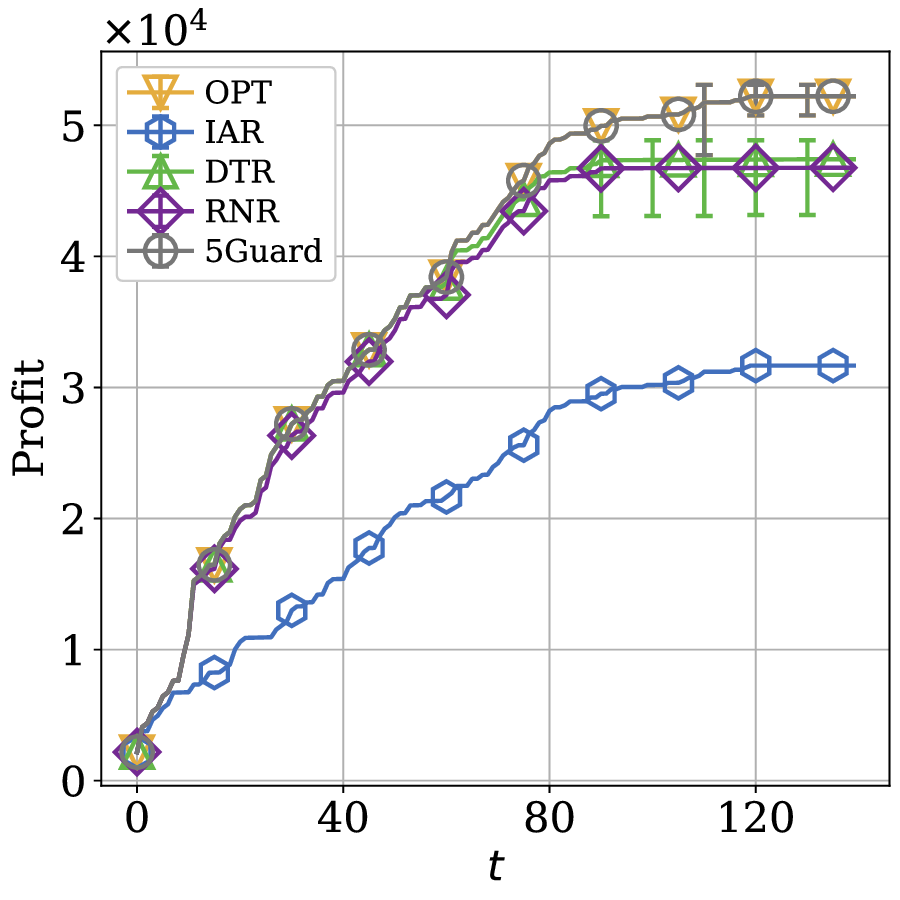}
        \label{figb-sim1}
    }
    \hfill
    \subfloat[]{%
        \includegraphics[width=0.24\textwidth]{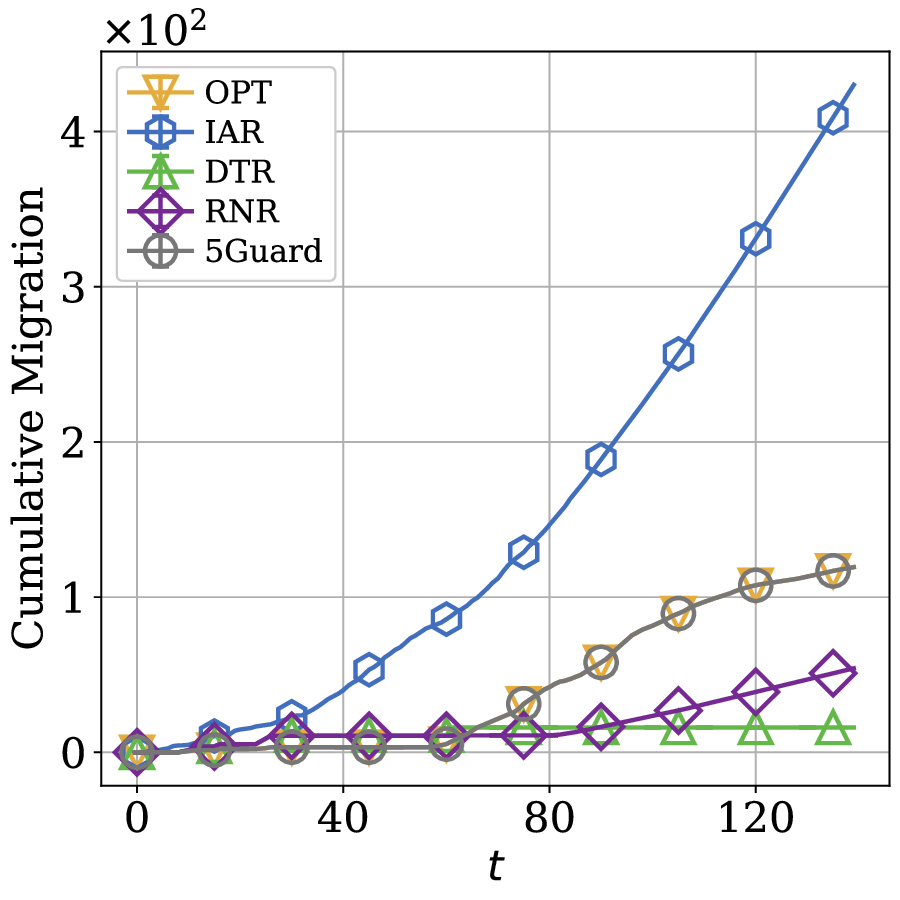}
        \label{figc-sim1}
    }
    \hfill
    \subfloat[]{%
        \includegraphics[width=0.24\textwidth]{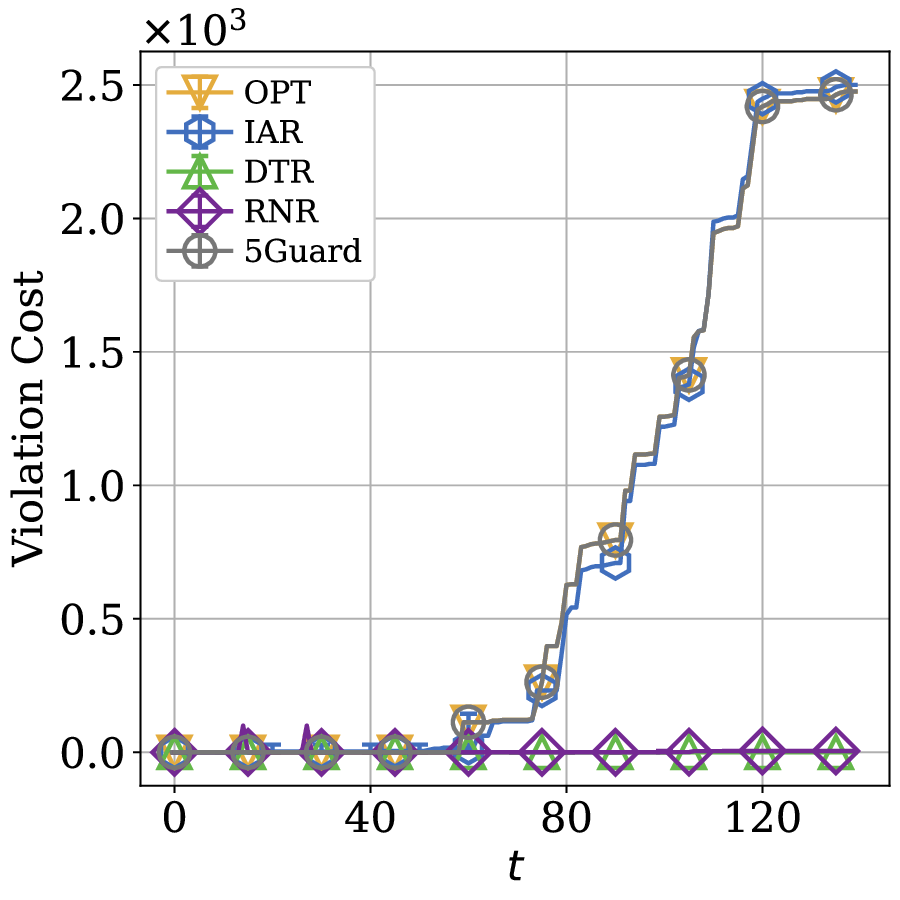}
        \label{figd-sim1}
    }
    \hfill
    \subfloat[]{%
        \includegraphics[width=0.24\textwidth]{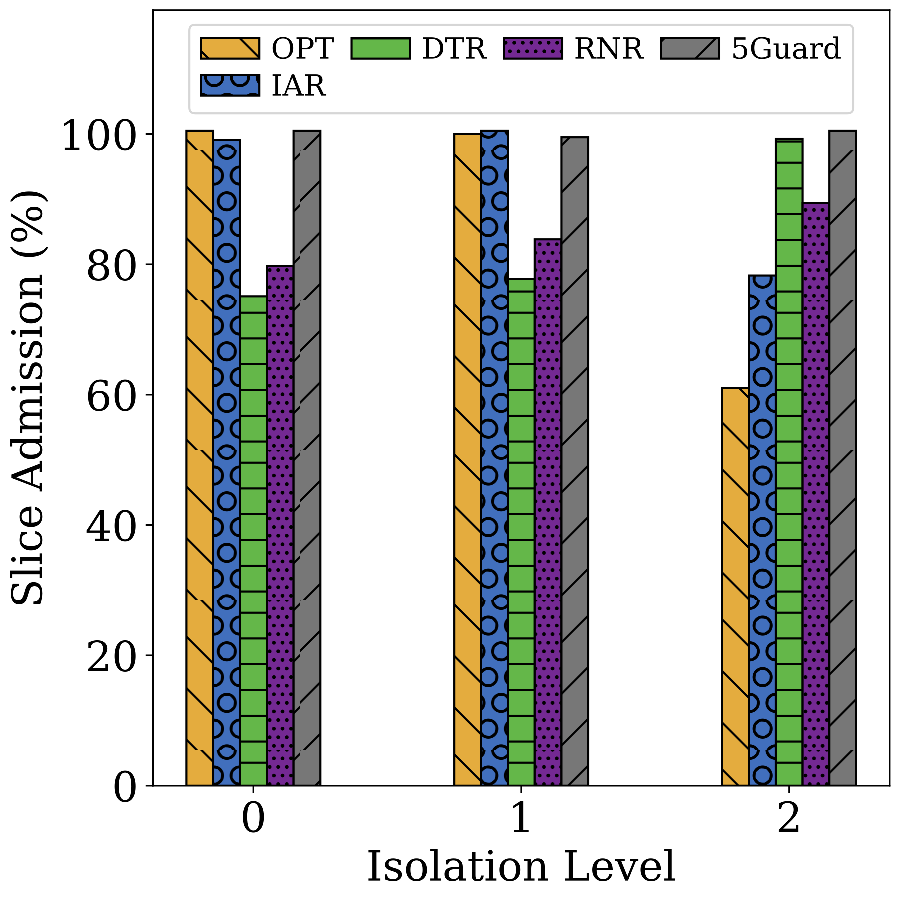}
        \label{fige-sim1}
    }
    \hfill
    \subfloat[]{%
        \includegraphics[width=0.24\textwidth]{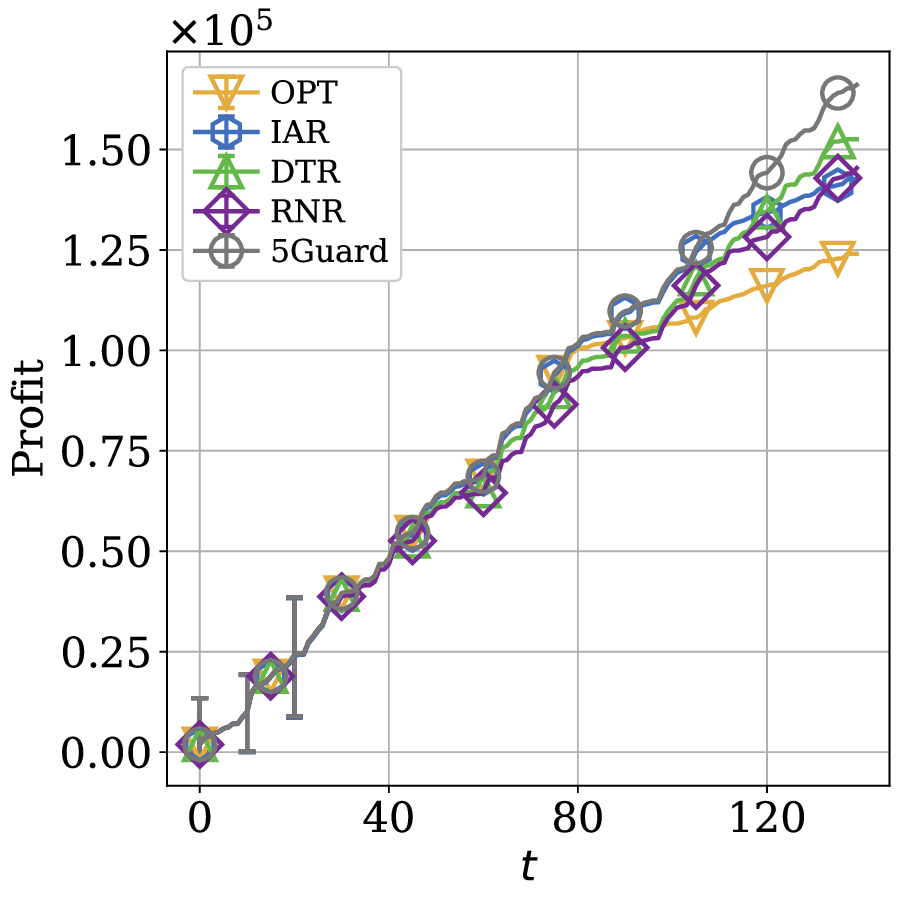}
        \label{figf-sim1}
    }
    \hfill
    \subfloat[]{%
        \includegraphics[width=0.24\textwidth]{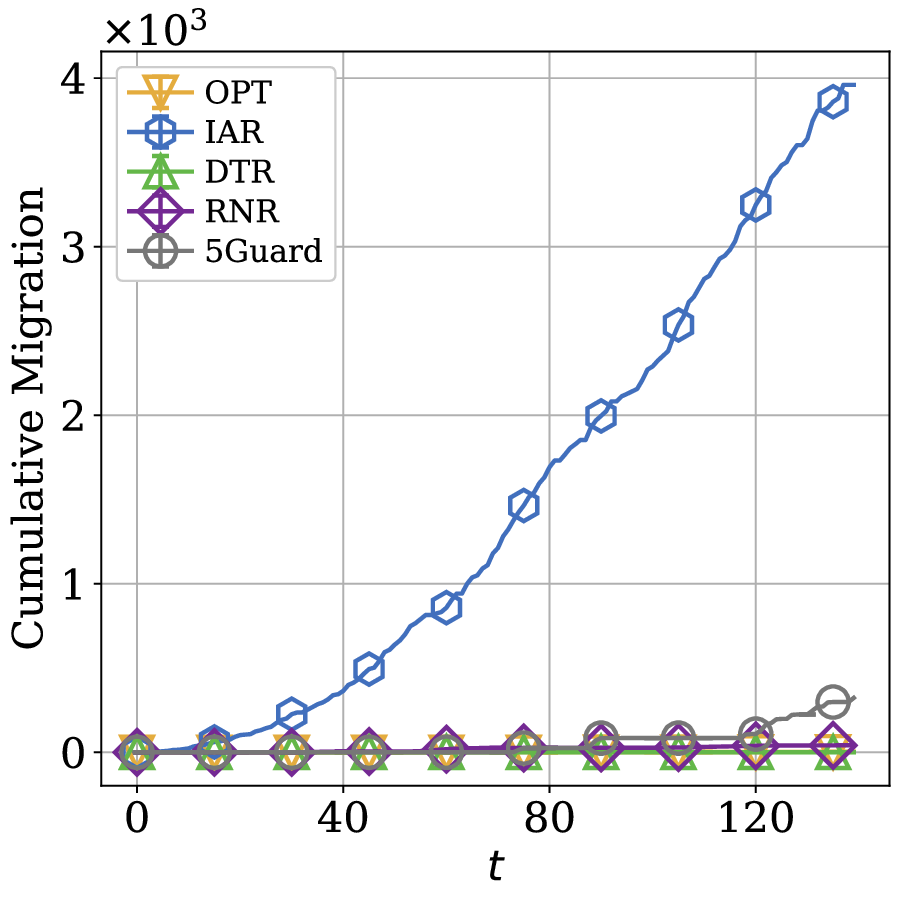}
        \label{figg-sim1}
    }
    \hfill
    \subfloat[]{%
        \includegraphics[width=0.24\textwidth]{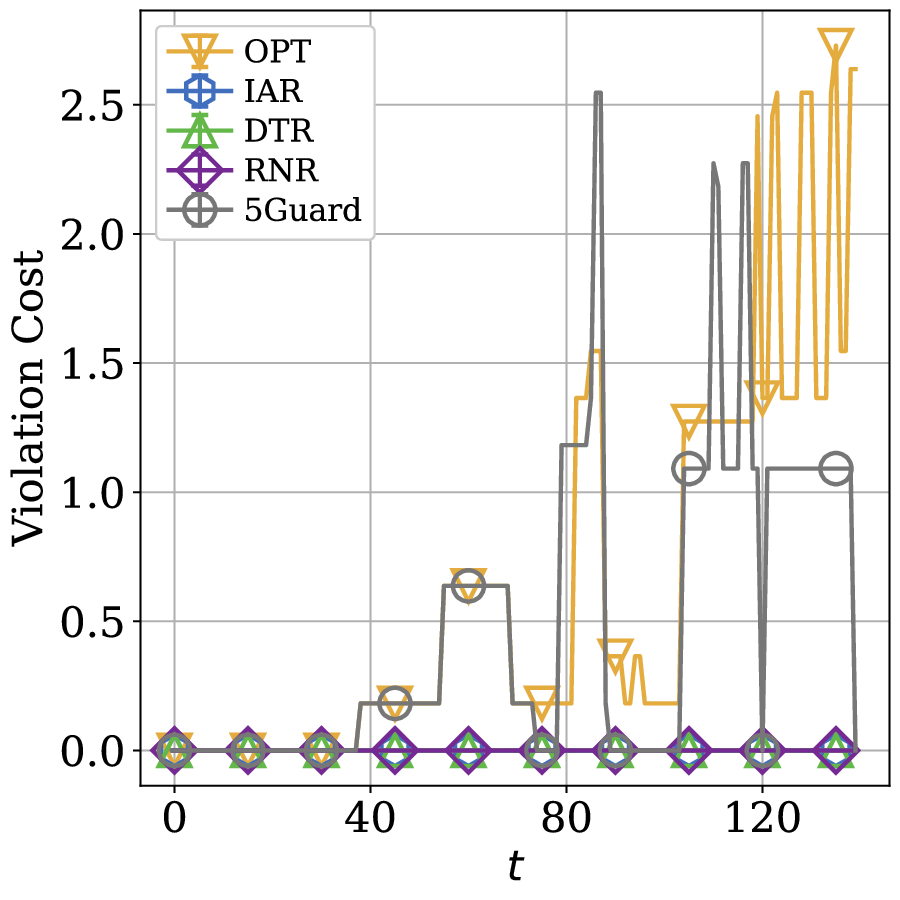}
        \label{figh-sim1}
    }

    \caption{Evaluation of SR admission percentage for each isolation level, profit, cumulative number of migrations, and violation cost over time steps by incrementally adding a new SR with different isolation levels at each time step for small-scale (i.e., a--d) and large-scale (i.e., e--h) 5G topologies.}
    \label{fig-sim1}
\end{figure*}
\begin{figure}
    \centering
    \includegraphics[width=0.8\linewidth]{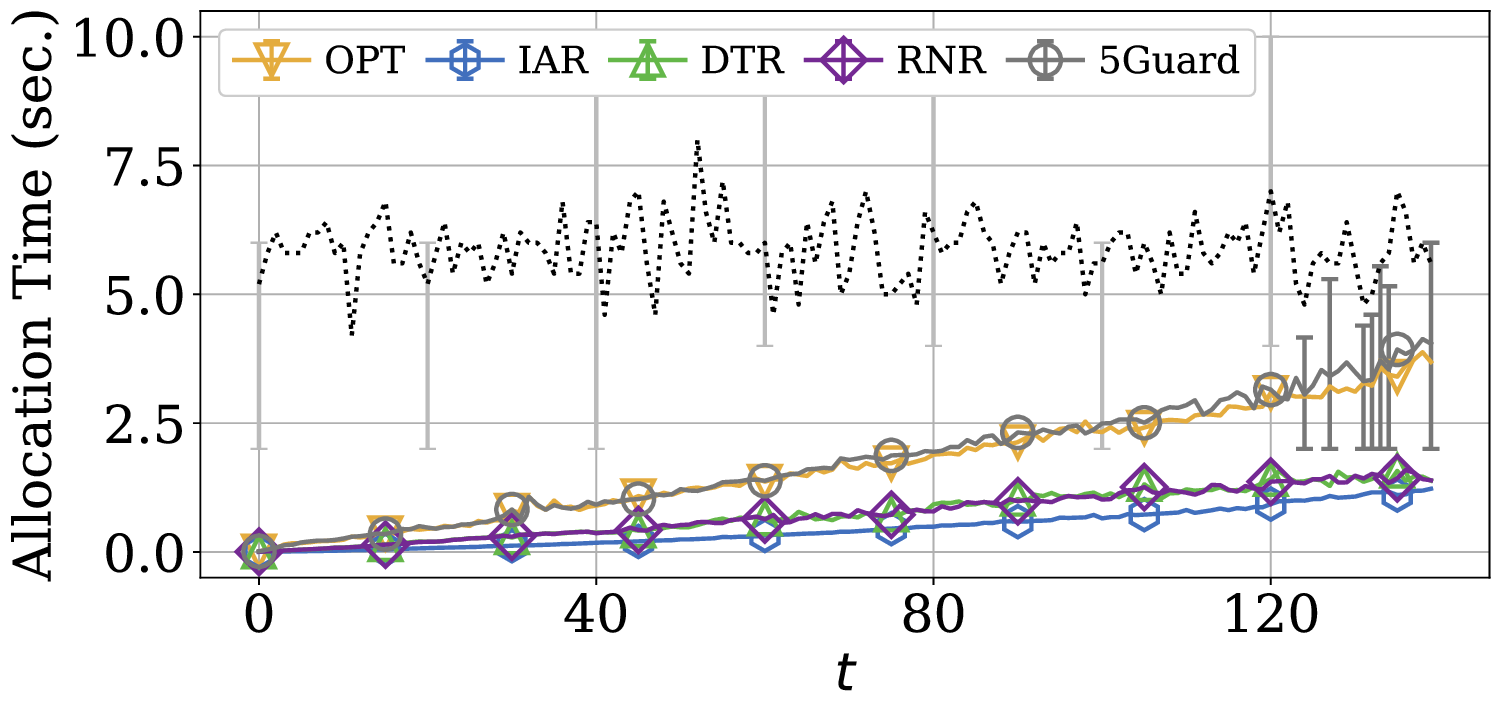}

    \caption{Resource allocation times with the BRAIN's 5G topology at each time step, with the dashed line indicating the average resource allocation deadlines of SRs alongside the error bars showing the range}
    \label{fig-sim2}
\end{figure}
\begin{figure}
    \centering
    \includegraphics[width=0.8\linewidth]{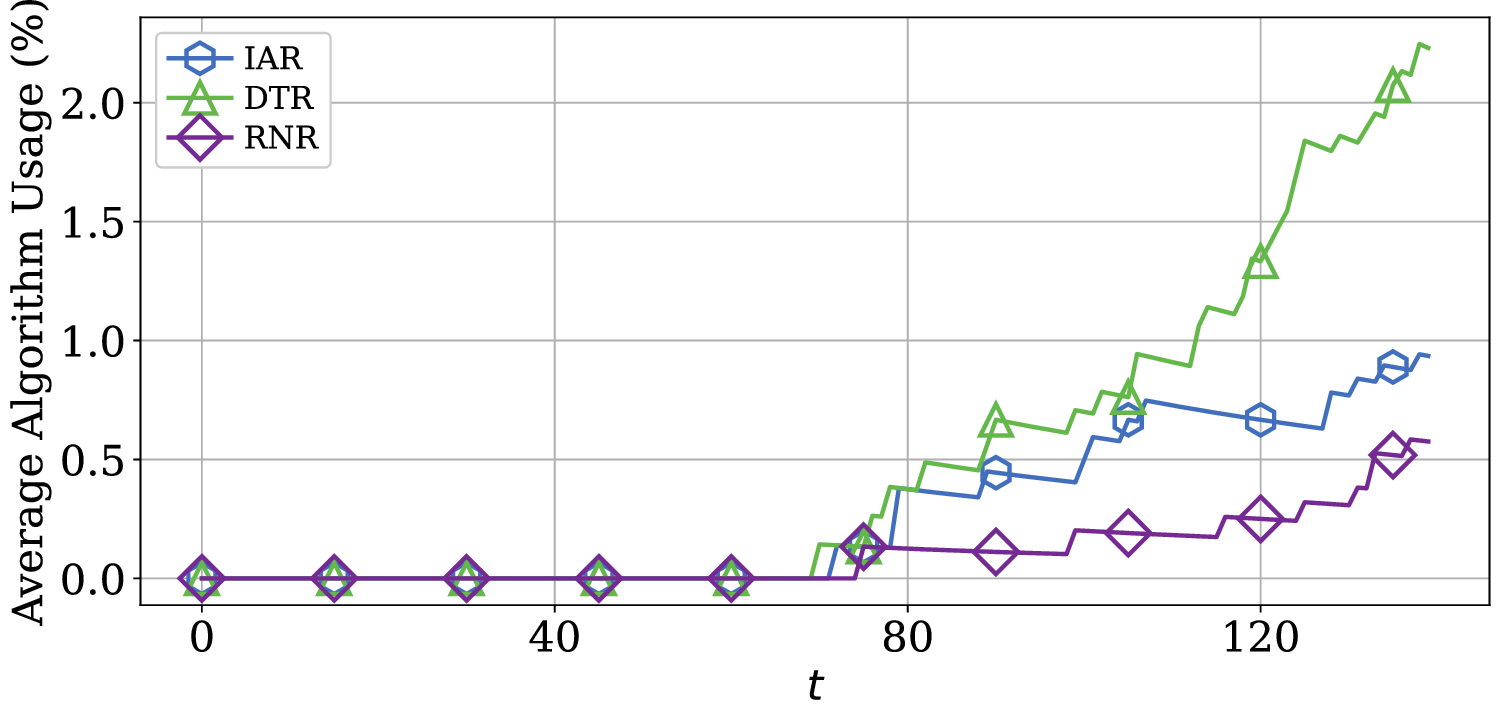}
	
    \caption{Running average percentage of usage for each non-OPT algorithm in 5Guard's ensemble while running on the BRAIN's 5G topology}
    \label{fig-sim3}
\end{figure}

We consider four types of slices: an L2 MC slice, an L1 uRLLC slice, an L0 mMTC slice without VNF sharing, and an L0 eMBB streaming slice with VNF sharing. To model a more realistic arrival process, SRs of these slices arrive randomly at each time step with probabilities of 0.7, 0.15, 0.1, and 0.05 for eMBB, uRLLC, mMTC, and MC slices, respectively. The results are averaged over multiple runs.

{\color{black}Fig. \ref{fig-srisol} shows the average isolation levels selected for arriving SRs at each time step $t$, with the shaded area indicating the range of selected isolation levels. Regarding resource allocation deadlines, since provisioning resources (e.g., K8s pods or VMs) typically takes at least seconds, deadlines should remain within the same order to avoid altering service access times. For our simulation, we set deadlines to 2 seconds for MC slices, 4 seconds for mMTC slices, 6 seconds for uRLLC slices, and 10 seconds for eMBB slices.}

{\color{black}Inspired by \cite{tmc25}, we adopt a pay-as-you-go revenue collection scheme, commonly used by InPs for service pricing. In our approach, pricing is based on deployment costs, which are significantly influenced by the isolation level imposed on the InP. These costs are then scaled by a coefficient that adjusts the profit, considering other factors, e.g., maintenance expenses.}
\begin{figure*}
    \centering
    \subfloat[]{%
        \includegraphics[width=0.24\textwidth]{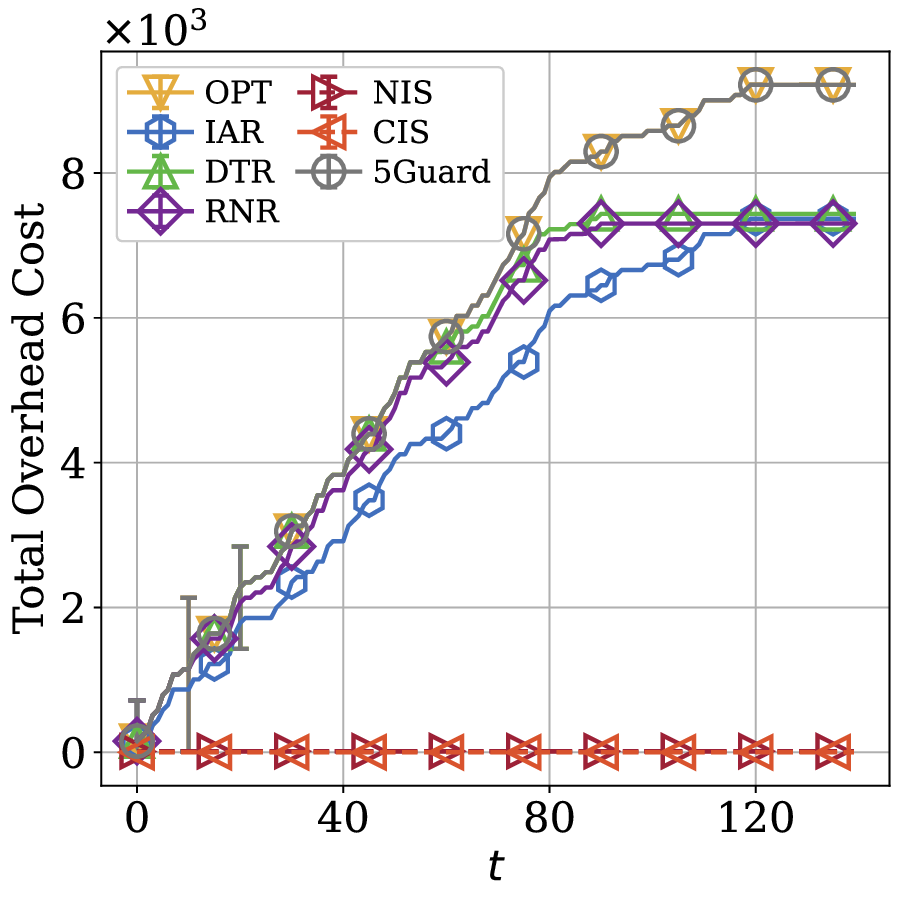}
        \label{figa-sim4}
    }
    \hfill
    \subfloat[]{%
        \includegraphics[width=0.24\textwidth]{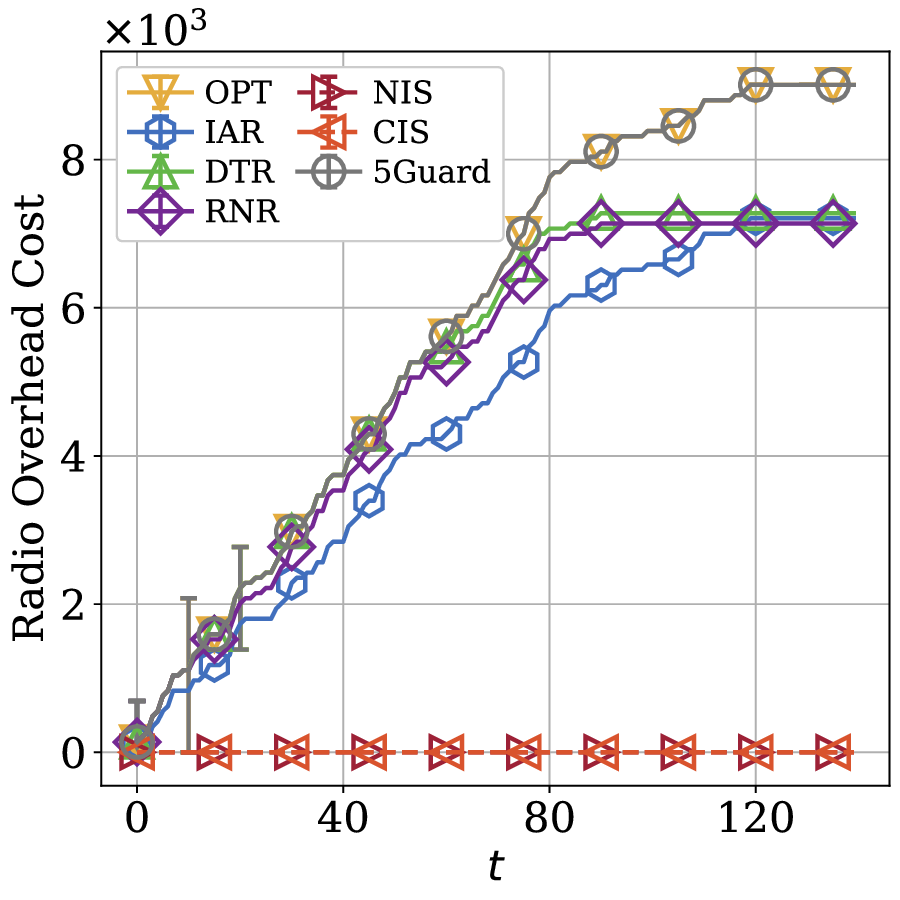}
        \label{figb-sim4}
    }
    \hfill
    \subfloat[]{%
        \includegraphics[width=0.24\textwidth]{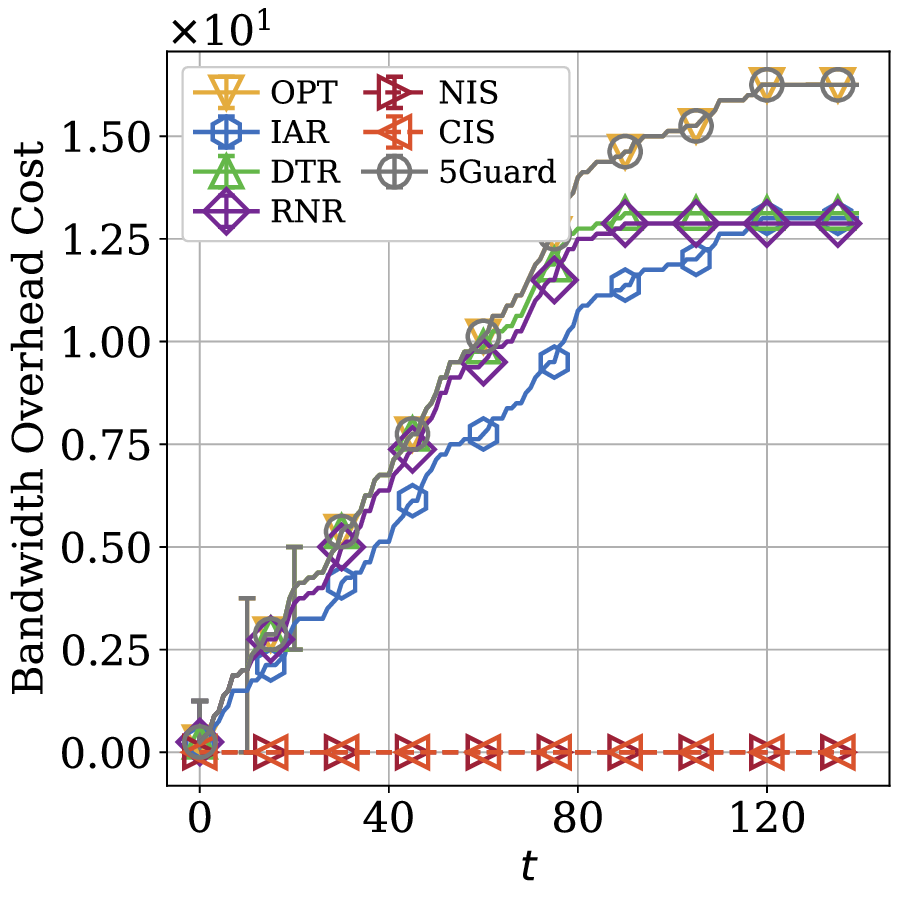}
        \label{figc-sim4}
    }
    \hfill
    \subfloat[]{%
        \includegraphics[width=0.24\textwidth]{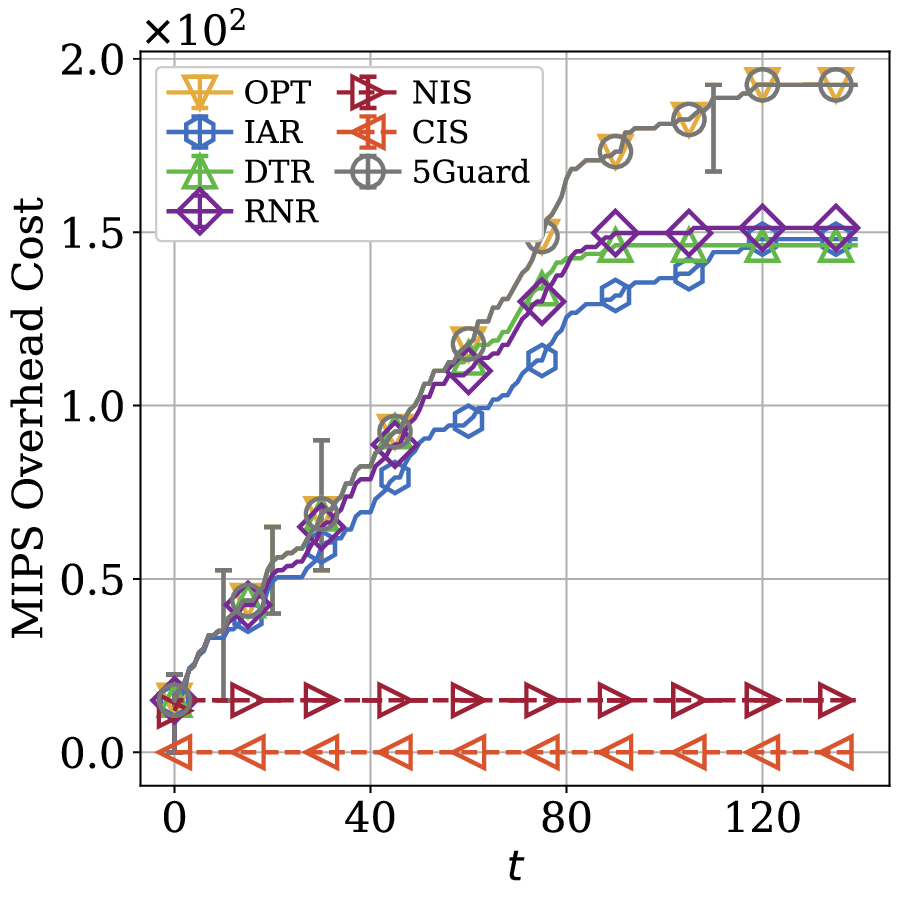}
        \label{figd-sim4}
    }
    \hfill
    \subfloat[]{%
        \includegraphics[width=0.24\textwidth]{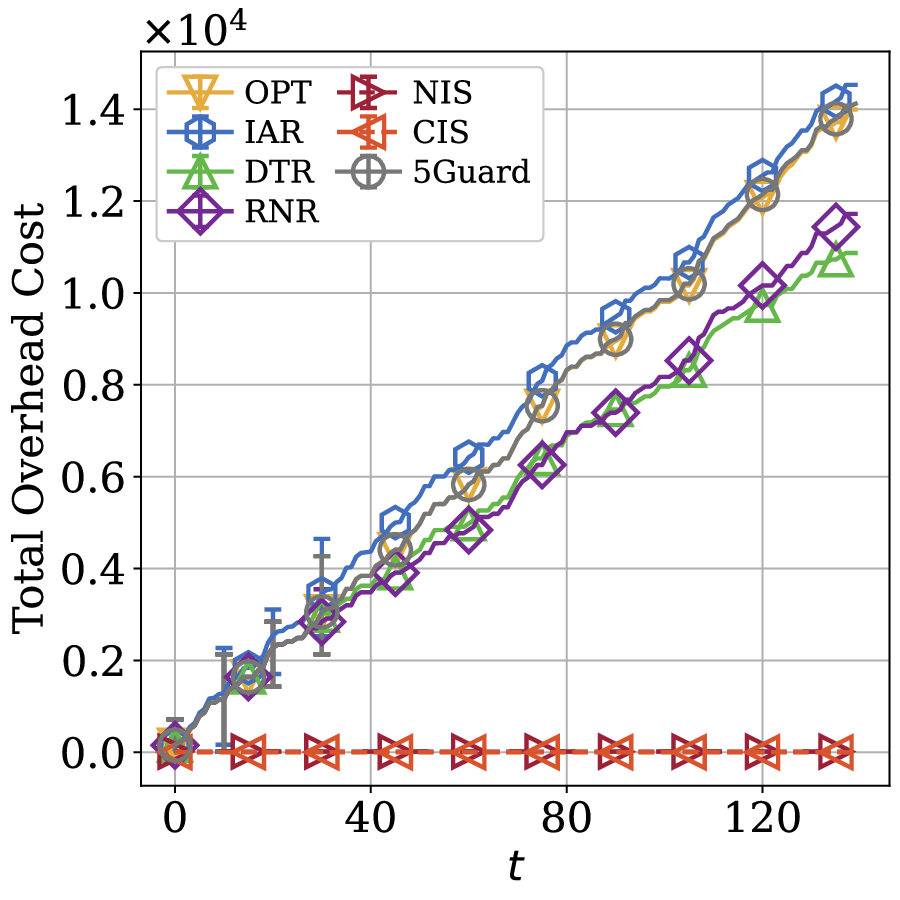}
        \label{fige-sim4}
    }
    \hfill
    \subfloat[]{%
        \includegraphics[width=0.24\textwidth]{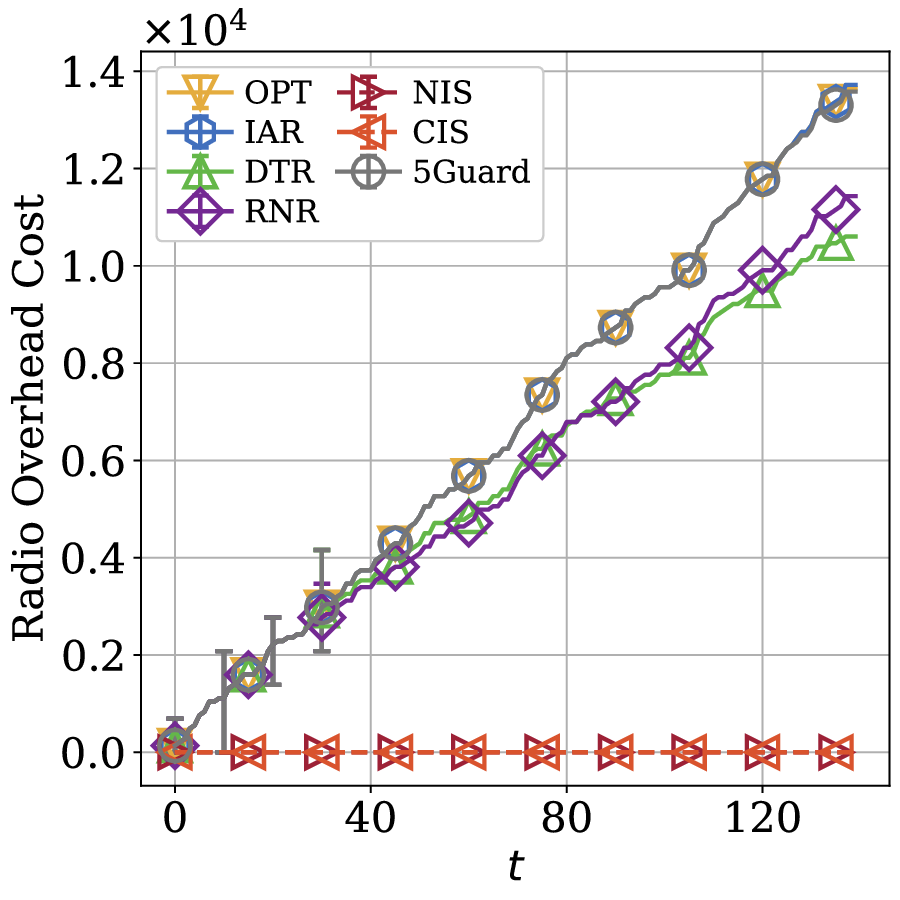}
        \label{figf-sim4}
    }
    \hfill
    \subfloat[]{%
        \includegraphics[width=0.24\textwidth]{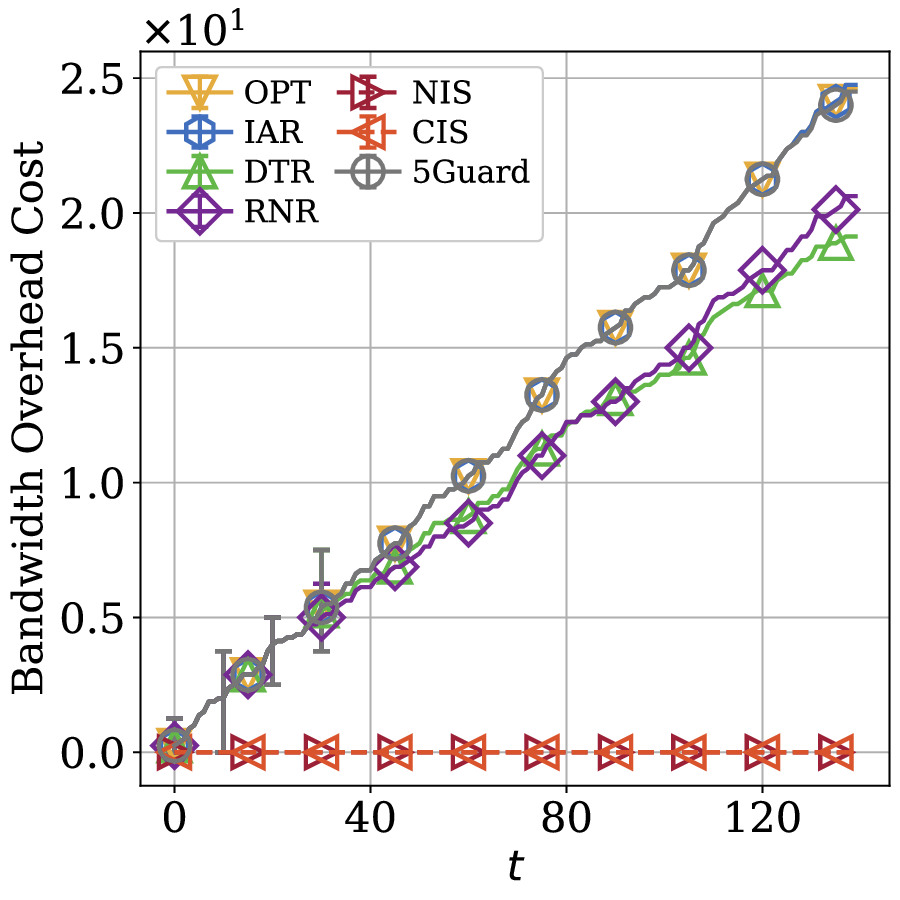}
        \label{figg-sim4}
    }
    \hfill
    \subfloat[]{%
        \includegraphics[width=0.24\textwidth]{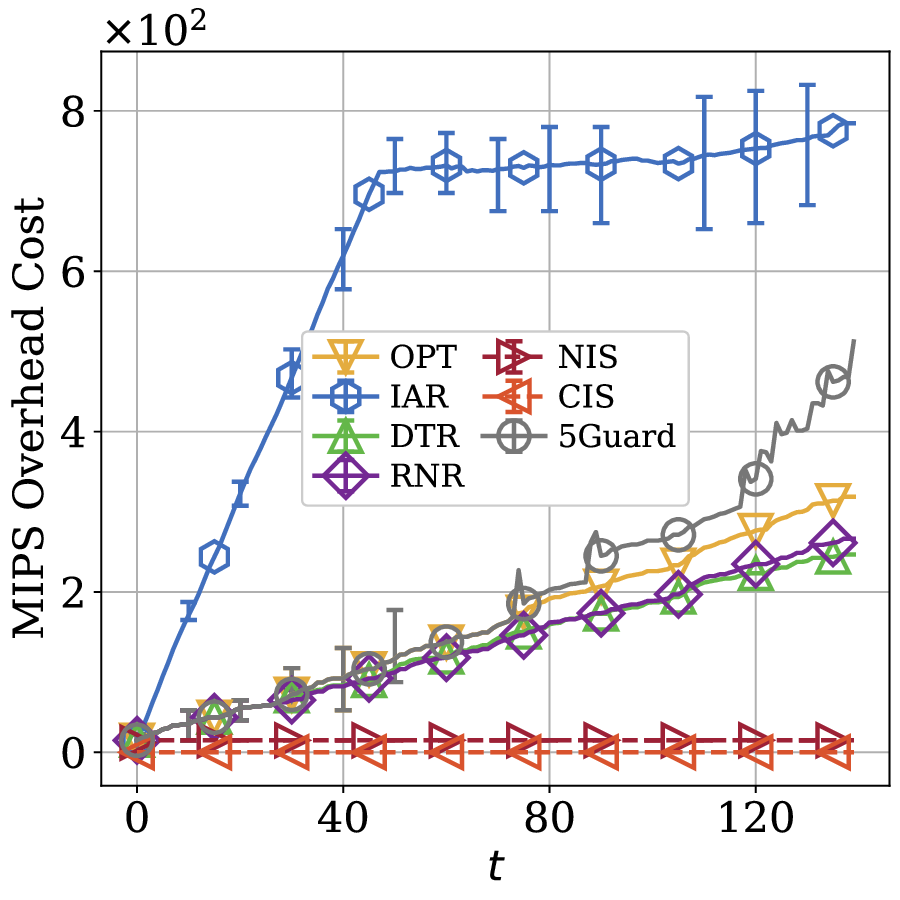}
        \label{figh-sim4}
    }
    \hfill
    \subfloat[]{%
        \includegraphics[width=0.24\textwidth]{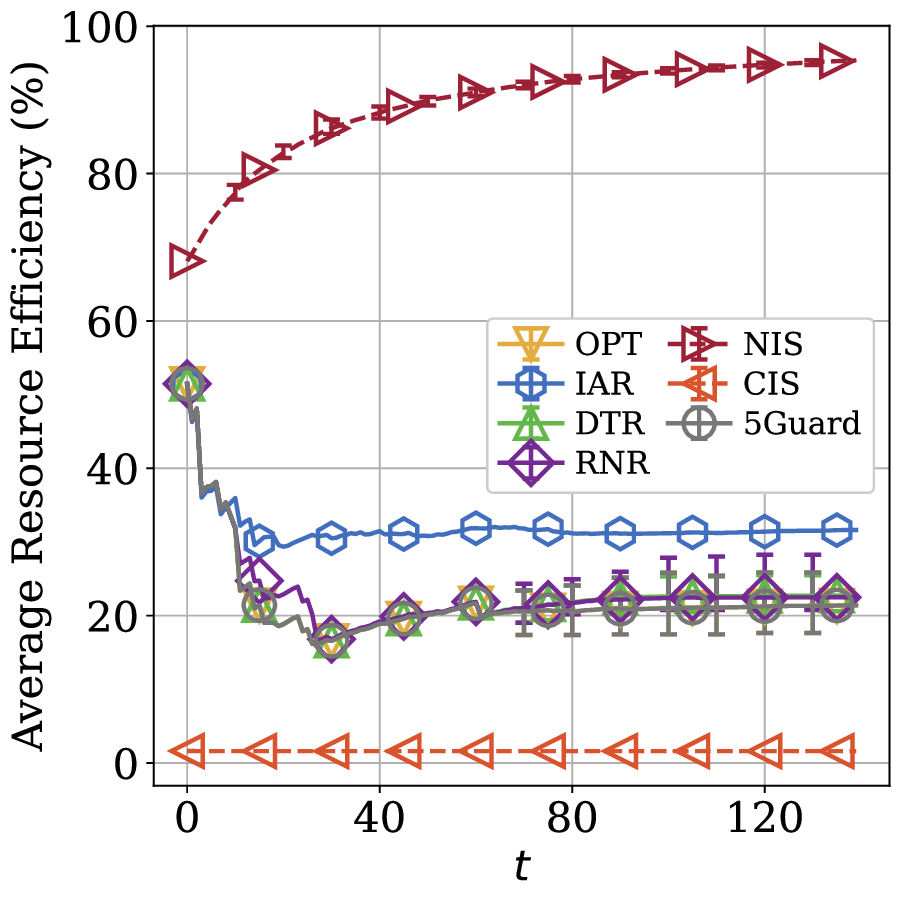}
        \label{figi-sim4}
    }
    \hfill
    \subfloat[]{%
        \includegraphics[width=0.24\textwidth]{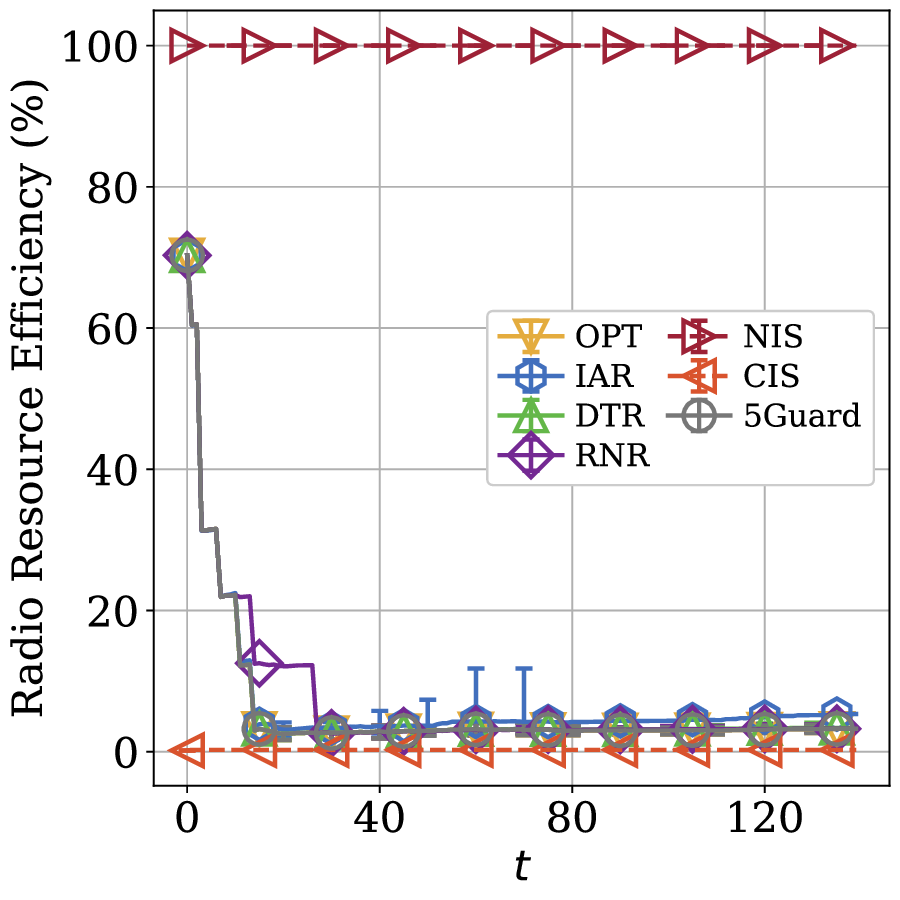}
        \label{figj-sim4}
    }
    \hfill
    \subfloat[]{%
        \includegraphics[width=0.24\textwidth]{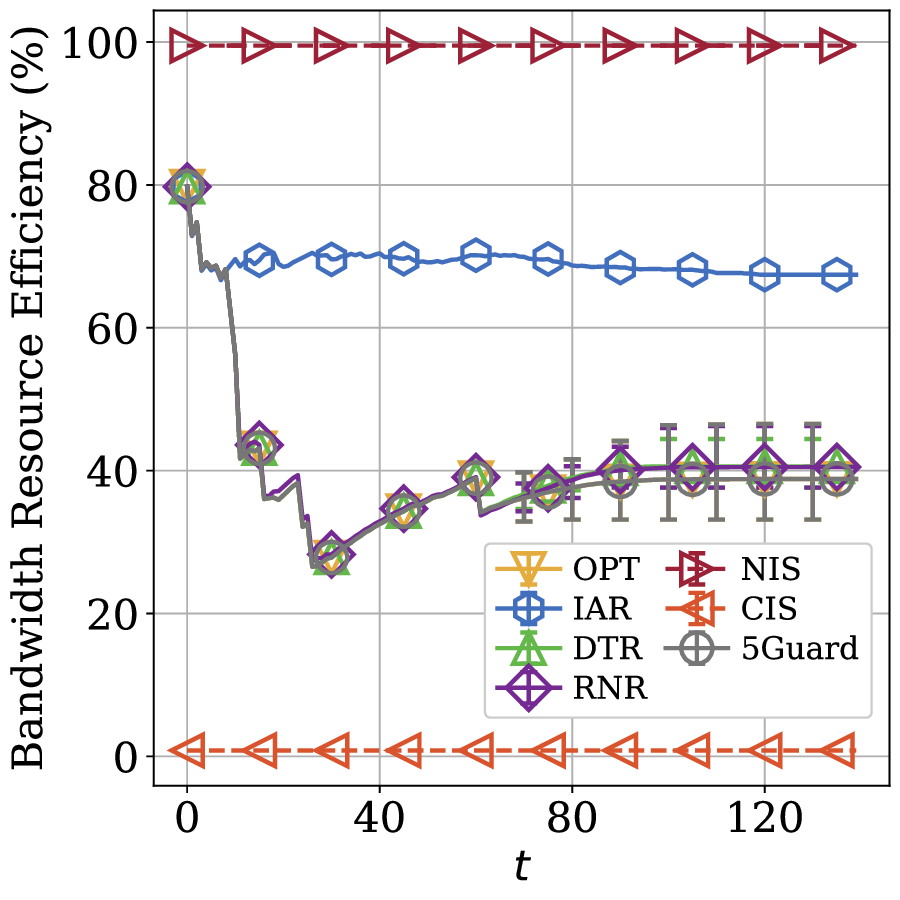}
        \label{figk-sim4}
    }
    \hfill
    \subfloat[]{%
        \includegraphics[width=0.24\textwidth]{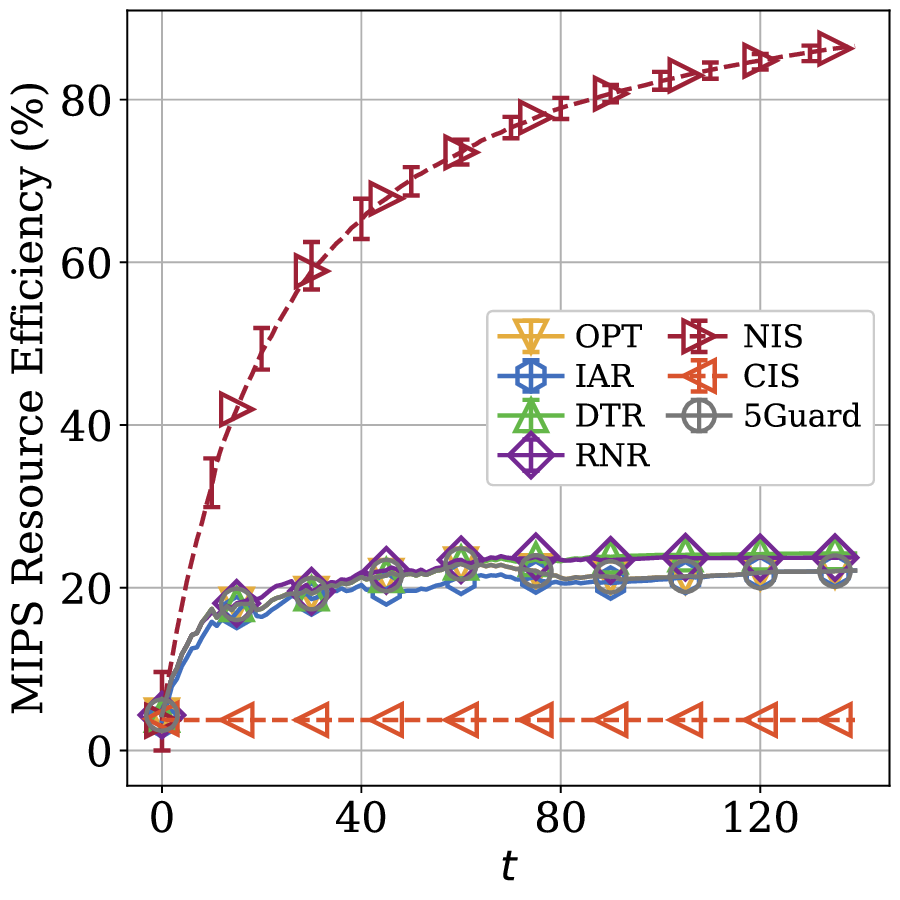}
        \label{figl-sim4}
    }
    \hfill
    \subfloat[]{%
        \includegraphics[width=0.24\textwidth]{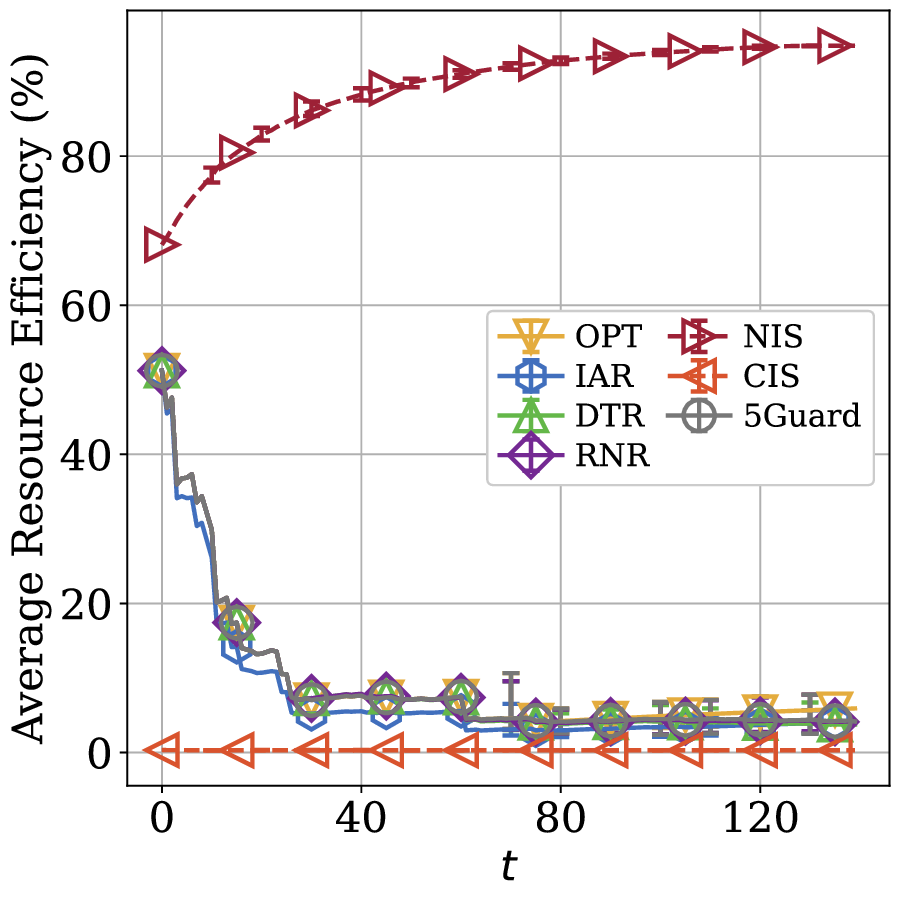}
        \label{figm-sim4}
    }
    \hfill
    \subfloat[]{%
        \includegraphics[width=0.24\textwidth]{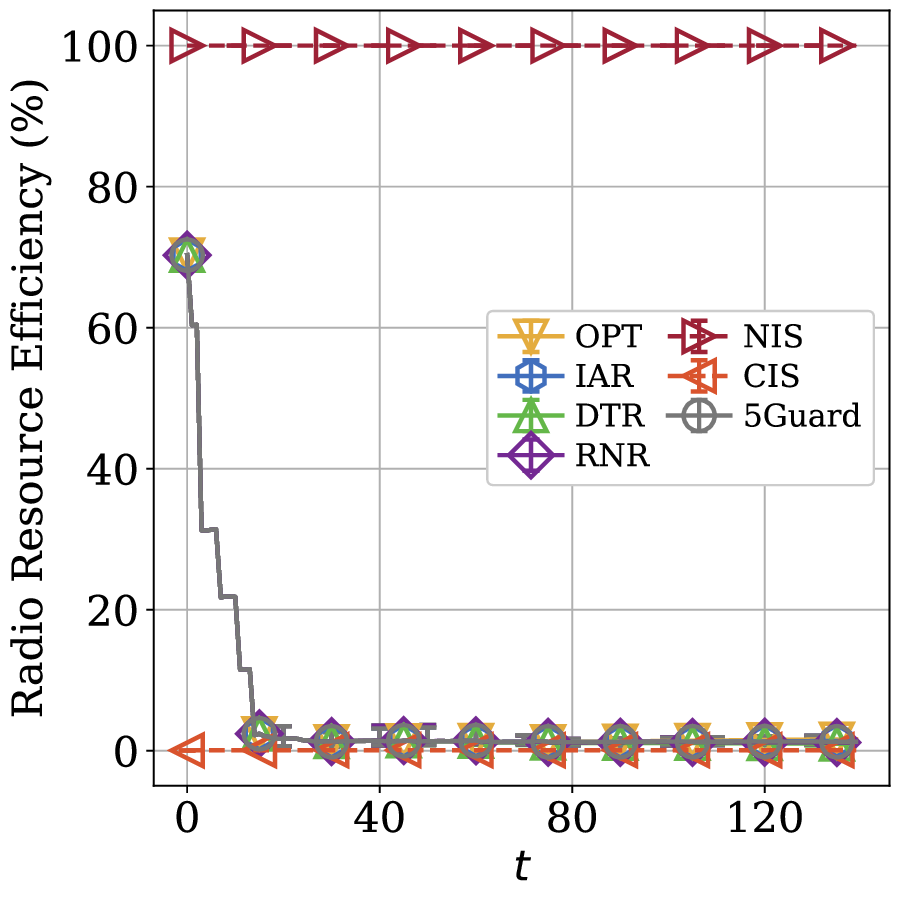}
        \label{fign-sim4}
    }
    \hfill
    \subfloat[]{%
        \includegraphics[width=0.24\textwidth]{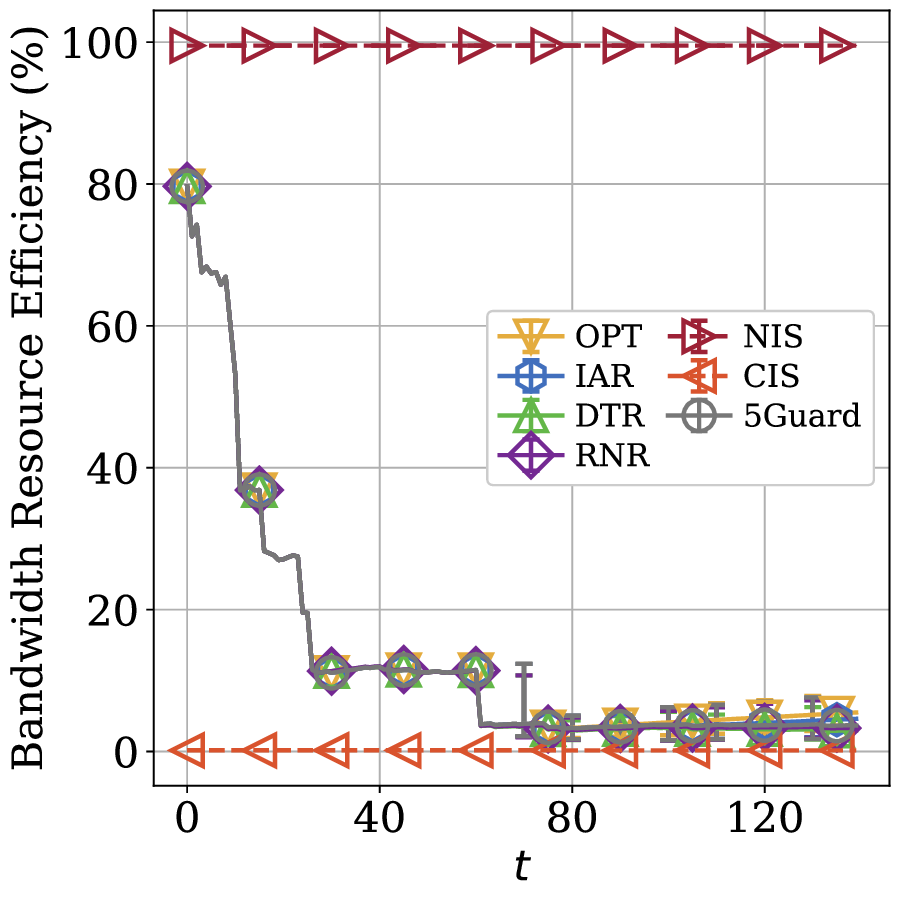}
        \label{figo-sim4}
    }
    \hfill
    \subfloat[]{%
        \includegraphics[width=0.24\textwidth]{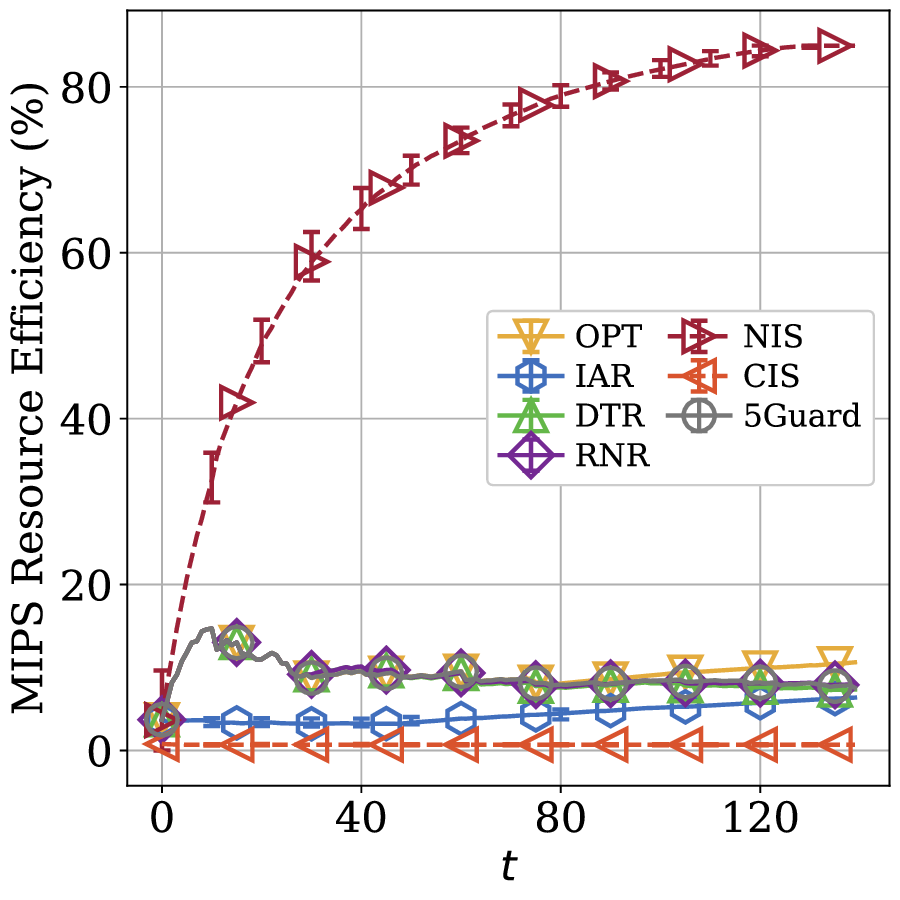}
        \label{figp-sim4}
    }

    \caption{Comparison of total overhead costs and domain-specific overhead costs along with average resource efficiency and domain-specific resource efficiency considering the isolation overheads and resource usage patterns at each time step for the synthetic (i.e., a--d and i--l) and BRAIN's 5G (i.e., e--h and m--p) topologies.}
    \label{fig-sim}
\end{figure*}
\subsection{Simulation Results}
\subsubsection{5Guard Evaluation}
Throughout the simulation runs, we modeled SR arrivals at each time step $t$. Figs. \ref{figa-sim1}--\ref{figd-sim1} present the SR admission percentages and profit, alongside the cumulative number of migrations and violation costs using the synthetic topology shown in Fig. \ref{figa-setup}. Migration and violation costs correspond to the two previously introduced techniques to increase admission.

{\color{black}As mentioned earlier, 5G-INS is the first isolation-aware formulation of 5G NS that integrates isolation levels according to standards. Consequently, comparing 5Guard with existing baselines that lack support for these levels may lead to an unfair evaluation due to differing SR inputs and, hence, different revenues under the pay-as-you-go scheme. According to Theorem 2, we proved that 5G-INS can be accurately transformed into a linear program with relaxed binary variables. Thus, OPT represents the optimal solution to 5G-INS, assuming all SRs are admitted whenever constraints allow it.}

As shown in Figs. \ref{figa-sim1}--\ref{figd-sim1}, 5Guard matches the performance of OPT. This is because the topology is small, allowing OPT to meet the deadlines for all SRs, including the L2 MC service for incoming SR $r_t\in\mathcal{R}$ at each time step $t$. Consequently, 5Guard achieved up to {\color{black}10.4}\% higher profit than DTR, the best-performing individual algorithm, and matched OPT. Since this topology includes only two NR-BS nodes, accepting more than one additional L2 SR is impossible after admitting one SR, as rejecting SRs post-admission is not allowed.

As depicted in Fig. \ref{figa-sim1}, 5Guard successfully admitted an L2 SR, achieving a non-zero admission rate, while allocating the remaining NR-BS/PS resources to L0/L1 SRs. Figs. \ref{figc-sim1}, \ref{figd-sim1}, \ref{figg-sim1}, and \ref{figh-sim1} illustrate how QoS violation and service migration strategies maximize profit. {\color{black}As stated earlier, the synthetic topology is designed to simulate a resource-constrained scenario. This demonstrates that controlled violation costs can significantly increase the admission of LP SRs in such conditions, where admission would otherwise be impossible.}

In Figs. \ref{fige-sim1}--\ref{figh-sim1}, we extended our investigation to the large-scale BRAIN's 5G topology shown in Fig. \ref{figb-setup}. 5Guard achieved up to 8.9\% more average profit than the best-performing algorithm in the ensemble and up to 33.89\% more average profit than OPT. It also achieved up to 25.4\% more admissions compared to the best-performing algorithm. The significant increase in input parameters caused a notable rise in OPT's execution time, occasionally leading to failures in meeting the resource allocation deadlines for MC and uRLLC SRs. {\color{black}However, we observed that running OPT for L0 SRs helped guide the allocation to superior solutions for later time steps, positively influencing the results even when switching to other algorithms.} Without loss of generality, we assumed resource allocation deadlines of 2 sec., 4 sec., 6 sec., and 10 sec. for MC, uRLLC, eMBB, and mMTC SRs, respectively. {\color{black}Note that these values differ from the delay requirements of SRs, which can be on the order of microseconds.}

According to Fig. \ref{fig-sim2}, from $t=70$, OPT failed to meet the MC SR requirements, and for $t>128$, it also could not meet the uRLLC SR requirements. {\color{black}Additionally, error bars of 5Guard’s allocation time for $t>70$ demonstrate ranges with minimum and maximum values set to 2 sec. and 4 sec., respectively. This behavior is due to 5Guard continuing the execution of algorithms up to the SR's deadline, as outlined in Algorithm \ref{alg1}. Furthermore, while OPT operates in a large-scale topology with many SRs, its allocation time does not show a drastic increase compared to other algorithms in the ensemble. However, based on our observations during the simulations, OPT can exhibit unreliability in specific scenarios, resulting in considerably larger allocation times. This does not imply that OPT is consistently unreliable or slow, especially in testbeds with many CPU cores allocated to the algorithms.

Figure~\ref{fig-sim3} illustrates the running average of usage percentages for non-OPT algorithms within the 5Guard framework. The results indicate that DTR is the most frequently selected algorithm, followed by IAR and RNR. Note that the percentage is controlled by how many L1 and L2 SRs we feed to the system, which have more stringent deadlines. This experiment helps determine which algorithms to include in the ensemble to minimize algorithm overhead while maximizing performance. As depicted in Fig. \ref{figf-sim1}, algorithms within the ensemble continuously surpass each other in profit, stemming from their distinct optimality-complexity trade-offs discussed earlier. As a result, all algorithms have been utilized by 5Guard, achieving 100\% admissions across the isolation levels.

Executing multiple algorithms concurrently can introduce additional overhead for the 5G InP. Based on our findings, in scenarios where the OPT algorithm consistently meets deadlines, the ensemble can be streamlined to include only OPT, thereby reducing computational complexity. Conversely, in environments characterized by expansive network topologies and stringent deadlines, excluding OPT from the ensemble may be advantageous due to its computational demands. In other situations, performance may vary over time, with different algorithms outperforming others depending on specific conditions. Therefore, incorporating diverse algorithms within the ensemble can enhance online performance by leveraging their complementary strengths. Developing adaptive strategies for ensemble management, which dynamically adjust the algorithm set based on real-time network conditions, presents a promising direction for future research.

Fig. \ref{figg-sim1} demonstrates that IAR caused significantly more migrations than the other algorithms. This is because it selects NR-BSs, PSs, and PPs with the maximum remaining resources such that when a new SR with higher revenue, isolation level, and priority than one or more pre-embedded SRs arrives, it can trigger the migration of all SRs with lower values.}

{\color{black}\subsubsection{Isolation Overhead Analysis}
In Figs. \ref{figa-sim4}--\ref{figh-sim4}, we present the overhead costs for different algorithms, with the two rows corresponding to the synthetic and BRAIN topologies, respectively. Additionally, we introduce two new baselines representing extreme cases of isolation-aware NS: NIS, which applies no isolation for all SRs, and CIS, which enforces complete isolation for all SRs. To compute overheads, we consider the resource portions highlighted in red in Fig. \ref{isolation}.

A key takeaway from these results is that, under NIS, the 5G InP experiences no overhead on radio resources and minimal overhead on bandwidth, but significantly higher overhead on MIPS resources. This is due to the absence of additional isolation guard bands for bandwidth and radio resources at L0. However, proper operation at L0 still requires extra headers for data transmission on VPs and additional infrastructure, such as OpenStack, a hypervisor, host and guest OSs, and a container orchestration tool for NFs.

Another observation is that overhead generally increases with the number of admitted SRs. However, MIPS overhead is also influenced by the placement of these SRs. OPT and 5Guard, which admit more SRs, also incur higher overhead costs. Notably, MIPS overhead spikes whenever a new VM or PS is used for the first time, which can significantly impact the 5G InP's profit at those time steps. Therefore, the InP should be carefully set $P_\min$ in (\ref{eq38}) to account for these fluctuations.

Additionally, IAR demonstrates significantly higher MIPS overhead in the BRAIN's 5G topology. This stems from IAR’s strategy of selecting NR-BS and PS nodes with the highest remaining resources, allocating a new, previously unused PS at each time step until all PSs are assigned at least one SR. The results show that the MIPS overhead has the steepest increase up to $t=51$, corresponding to the number of PSs in the BRAIN topology.

Another insight is that radio overhead cost is the primary contributor to total overhead cost. This is particularly evident when the NR-BS bandwidth is 50 MHz, translating to approximately 2778 PRBs. Given the resource requirements, this can become a bottleneck, leading to a higher cost per resource unit and, consequently, a higher overall overhead cost.

The other aspect of isolation we study is its impact on resource usage patterns, as discussed earlier. Figs. \ref{figi-sim4}--\ref{figp-sim4} illustrate the average and domain-specific resource efficiency for the synthetic and the BRAIN's 5G topologies. These figures account for the overhead costs and the resource usage patterns mentioned previously.

As expected, enforcing complete isolation results in the lowest possible resource efficiency, whereas the absence of isolation yields the highest efficiency. For MIPS resources, efficiency starts at a low value and gradually increases as more SRs are admitted. This is because, initially, a significant portion of the consumed MIPS is allocated to overheads such as the hypervisor, OpenStack, and host and guest OSs, rather than the actual hosting of NFs. However, under NIS, MIPS efficiency recovers more rapidly due to the ability to share NFs and instantiate fewer containers or VM instances, ultimately leading to higher resource efficiency. 

Importantly, isolation-aware approaches enable us to meet SLA requirements while achieving significantly greater efficiency than CIS, particularly for bandwidth and MIPS resources. The difference between isolation-aware methods and CIS for radio resources becomes more evident in scenarios where NR-BS bandwidth is higher.

In contrast to the synthetic topology, where a maximum of only two L2 SRs could be admitted, the BRAIN's 5G topology allows for a significantly higher number of L2 SR admissions. This leads to resource efficiency percentages converging more closely to CIS, especially for radio resources.

\subsection{Key Remarks and Limitations}
As this work aimed at studying the impact of incorporating isolation levels into 5G NS, 5Guard operates under the assumption that the InP admits all SRs whenever the constraints permit. Consequently, 5Guard does not include a proactive admission control mechanism. Admitting all SRs can influence the overall admission percentage and the long-term profit of the 5G InP. However, a critical remark about 5Guard is that, since it maximizes online profit at each time step, it can be well integrated with reward-based admission control schemes, where the 5Guard output serves as a reward signal.
}

\section{Conclusion and Future Work}
This paper explored the theoretical and practical implications of incorporating isolation into 5G network slicing. We formulated the isolation-aware 5G network slicing problem and studied its complexity. We then proposed a novel adaptive framework, 5Guard, which solves the problem by leveraging an ensemble of algorithms ranging from fast to exact methods, selected based on resource allocation deadlines. Next, we evaluated 5Guard's efficacy in challenging scenarios and large-scale networks through simulation. Finally, we analyzed the impact of isolation on resource overhead and efficiency for different algorithms.

For future work, we will investigate the impact of sequential learning-based admission control on the long-term performance of 5Guard. Another direction is to design a block that intelligently selects the appropriate set of algorithms for the ensemble in 5Guard. Finally, we aim to develop solutions for translating security requirements into isolation levels.


\begin{appendices}
    \section{Proof of Theorem 1}\label{appa}
    We prove Theorem 1 by contradiction. Thus, we first assume that 5G-INS does not belong to the set of NP-hard problems. Consider a subset of instances of 5G-INS with a clean slate on the PN and $r_t$ as the incoming HP slice request with a given isolation level and QoS requirements. Under these assumptions, we can eliminate the binary variable vector $\boldsymbol{y}$ by merging it with $\boldsymbol{x}$ and $\boldsymbol{z}$ since resources are not fragmented in a clean slate. We can also eliminate capacity and PRB allocation variables $\boldsymbol{c}$ and $\boldsymbol{n}$, setting their components with $x_v^pC_v^{\text{REQ}}$. We then assume there is at most a single PP with the least cost between two individual PS/NR-BS nodes, eliminating bandwidth allocation variables $\boldsymbol{b}$ and further simplifying decision-making. The bandwidth can then be set as $B_{vw}^{\text{REQ}}x_v^px_w^p$.

    We know that in 5G-INS, each NF/RU maps to a single PS/NR-BS, and we aim to minimize mapping costs given a revenue. Specifically, we create a bin-packing problem where items (i.e., NF/RU nodes) have sizes $C_v^{\text{REQ}}$ for NFs and $K_w^{\text{REQ}}$ for RUs. Bins (i.e., PS/NR-BS nodes) have capacities $C_p^{\max}$ for each NF $p\in\mathcal{V}^{\text{P,S}}$ and $\Pi_q^{\max}$ for each RU $q\in\mathcal{V}^{\text{P,R}}$, and the cost of assigning an item to a bin is set to 1. With this definition, any solution to this bin-packing problem can be mapped to a solution of the simplified 5G-INS problem. This establishes that the simplified problem is NP-hard since bin-packing is a known NP-hard problem \cite{whf79}. As this is a subset of 5G-INS instances and is proven NP-hard, then solving 5G-INS must involve solving instances that are at least as hard as this simplified NP-hard instance \cite{whf79}. Therefore, 5G-INS inherits the hardness of its NP-hard subset. This contradicts the assumption of 5G-INS not being in the set of NP-hard problems and completes the proof.

    \section{Proof of Theorem 2}\label{appb}
    By relaxing $\boldsymbol{x}$, $\boldsymbol{y}$, and $\boldsymbol{z}$, constraints (\ref{eq1}), (\ref{eq2}), (\ref{eq8}), (\ref{eq9}), (\ref{eq23}), (\ref{eq25}), and (\ref{eq26}) become affine sets with respect to their corresponding variables. In constraints (\ref{eq4})--(\ref{eq7}), (\ref{eq15}), (\ref{eq16}), (\ref{eq21}), (\ref{eq22}), (\ref{eq28}), and (\ref{eq38}), we encounter terms in the form of $g^{\text{aux}}(\boldsymbol{\theta})\triangleq\max(g^{\text{aux}}_1(\boldsymbol{\theta}),...,g^{\text{aux}}_n(\boldsymbol{\theta}))$, where $\boldsymbol{\theta}$ is a placeholder for a vector of binary variables and $g^{\text{aux}}_i(\boldsymbol{\theta}), i\in[1,n]$ are linear functions with binary outputs. {\color{black}Note that $\min(\cdot,\cdot)$ functions in constraint (\ref{eq28}) can be trivially rewritten with $\max(\cdot,\cdot)$}. With relaxation, we know that these point-wise $\max$ functions become convex, since $g^{\text{aux}}_i(\boldsymbol{\theta})$ functions are linear and hence convex. However, we can convert them to affine constraints, which can be more efficient for use with methods such as interior point. To accomplish this, we first define an auxiliary relaxed binary variable $\hat{u}$ with 
    \begin{equation}\label{eq40}
        \begin{aligned}
            \hat{u}\geq g^{\text{aux}}_i(\boldsymbol{\theta}),\quad\forall i\in[1,n]
        \end{aligned}
    \end{equation}
    \begin{equation}\label{eq41}
        \begin{aligned}
            \hat{u}\leq \sum_{i=1}^ng^{\text{aux}}_i(\boldsymbol{\theta})
        \end{aligned}
    \end{equation}
    Constraint set (\ref{eq40}) enforces $u\geq g^{\text{aux}}(\boldsymbol{\theta})$, which is the epigraph of $g^{\text{aux}}(\boldsymbol{\theta})$. To extract $g^{\text{aux}}(\boldsymbol{\theta})$, we need to cover the case with $g^{\text{aux}}_i(\boldsymbol{\theta})=0, \forall i\in[1,n]$, that is achieved by (\ref{eq41}). For relaxed $u$, (\ref{eq4})--(\ref{eq7}), (\ref{eq21}), (\ref{eq22}), (\ref{eq28}), (\ref{eq40}), and (\ref{eq41}), become affine sets.
    
    In (\ref{eq33}), we have $\min$ and $\max$ functions. As $\sum_{\lambda\in\mathcal{D}_l}z_{vw}^{l,\lambda}$ outputs binary values{\color{black}, similar to (\ref{eq40}) and (\ref{eq41}),} we can define an auxiliary variable $\Tilde{u}_{vw}^{pqi}$ for $\min_{l\in\mathcal{L}_{pqi}}\sum_{\lambda\in\mathcal{D}_l}z_{vw}^{l,\lambda}$ with the following constraints.
    \begin{displaymath}
        \Tilde{u}_{vw}^{pqi}\leq \sum_{\lambda\in\mathcal{D}_l}z_{vw}^{l,\lambda},\quad\forall l\in\mathcal{L}_{pqi},  e_{pq}^{(i)}\in\mathcal{E}^{\text{P}},
         e_{vw}\in\mathcal{V}_r,  r\in\mathcal{R}
    \end{displaymath}
    \begin{displaymath}
            \Tilde{u}_{vw}^{pqi}\geq \sum_{l\in\mathcal{L}_{pqi}}\sum_{\lambda\in\mathcal{D}_l}z_{vw}^{l,\lambda} - |\mathcal{L}_{pqi}| + 1,~ 
            \text{\footnotesize$\forall e_{pq}^{(i)}\in\mathcal{E}^{\text{P}}, e_{vw}\in\mathcal{V}_r,  r\in\mathcal{R}$}
    \end{displaymath}
    where the latter {\color{black}ensures $\Tilde{u}_{vw}^{pqi}=1$ when $\sum_{\lambda\in\mathcal{D}_l}z_{vw}^{l,\lambda}=1, \forall l\in\mathcal{L}_{pqi}$, by creating two constraints $\Tilde{u}_{vw}^{pqi}\leq1$ and $\Tilde{u}_{vw}^{pqi}\geq1$ to resemble the $\min$ function mentioned. In addtion, $\Tilde{u}_{vw}^{pqi}$ can be reused for (\ref{eq20}).}
    
    {\color{black}We also have another $\max$ function in (\ref{eq33}). Note that $[\cdot]^+\triangleq\max(0,\cdot)$. To convert this,} we need two sets of auxiliary variables $\hat{a}_{vw}$ and $\Tilde{a}_{vw}$, and the following constraints.
    \begin{displaymath}
        \hat{a}_{vw}\geq \sum_{e_{pq}^{(i)}\in\mathcal{E}^{\text{P }}}D_{pqi}^{\text{RTT}}\Tilde{u}_{vw}^{pqi}-h_{vw}^{\text{D}},\quad\forall e_{vw}\in\mathcal{V}_r,  r\in\mathcal{R}
    \end{displaymath}
    \begin{displaymath}
        \hat{a}_{vw}\geq 0,\quad\forall e_{vw}\in\mathcal{V}_r, r\in\mathcal{R}
    \end{displaymath}
    \begin{displaymath}
        \hat{a}_{vw}\leq \sum_{e_{pq}^{(i)}\in\mathcal{E}^{\text{P }}}D_{pqi}^{\text{RTT}}(1-\Tilde{a}_{vw}),\quad\forall e_{pq}^{(i)}\in\mathcal{E}^{\text{P}},e_{vw}\in\mathcal{V}_r,  r\in\mathcal{R}
    \end{displaymath}
    \begin{displaymath}
        \begin{aligned}
    \hat{a}_{vw}\leq \sum_{e_{pq}^{(i)}\in\mathcal{E}^{\text{P }}}D_{pqi}^{\text{RTT}}\Tilde{u}_{vw}^{pqi}-h_{vw}^{\text{D}} + \sum_{e_{pq}^{(i)}\in\mathcal{E}^{\text{P }}}D_{pqi}^{\text{RTT}}\Tilde{a}_{vw},&\\
            \forall e_{pq}^{(i)}\in\mathcal{E}^{\text{P}}, e_{vw}\in\mathcal{V}_r,  r\in\mathcal{R}&
        \end{aligned}
    \end{displaymath}
    where $\Tilde{a}_{vw}$ is a binary variable. {\color{black}The first two constraints create the epigraph similar to (\ref{eq40}). However, since the terms inside $\max$ are not binary, we need extra constraints and variables to make them equivalent. The other two constraints accomplish this. If revisitation occurs, i.e. $\Tilde{h}_{vw}^{\text{REV}}=0$, we should have $\hat{a}_{vw}=0$ and otherwise the term enclosed by $[\cdot]^+$ in (\ref{eq33}). After setting $M= \sum_{e_{pq}^{(i)}\in\mathcal{E}^{\text{P }}}D_{pqi}^{\text{RTT}}$ in $h_{vw}^{\text{D}}$, we defined $\Tilde{a}_{vw}$ as a binary variable which creates the following two cases:
    \begin{itemize}
        \item Revisitation occurs, i.e., $\Tilde{h}_{vw}^{\text{REV}}=0$:\\
            $\hat{a}_{vw}\leq \sum_{e_{pq}^{(i)}\in\mathcal{E}^{\text{P }}}D_{pqi}^{\text{RTT}}(1-\Tilde{a}_{vw})$,\\
            $\hat{a}_{vw}\leq \sum_{e_{pq}^{(i)}\in\mathcal{E}^{\text{P }}}D_{pqi}^{\text{RTT}}\Tilde{u}_{vw}^{pqi}-(\sum_{e_{pq}^{(i)}\in\mathcal{E}^{\text{P }}}D_{pqi}^{\text{RTT}}+\sum_{e_{pq}^{(i)}\in\mathcal{E}^{\text{P }}}D_{vw}^{\text{REQ}}) + \sum_{e_{pq}^{(i)}\in\mathcal{E}^{\text{P }}}D_{pqi}^{\text{RTT}}\Tilde{a}_{vw}$\\
            In this case, $\Tilde{a}_{vw}$ will be forced to 1 since the other case can lead to an infeasible case. Then, $\hat{a}_{vw}=0$ as desired since we will have two constraints in the form of $\hat{a}_{vw}\geq0$ and $\hat{a}_{vw}\leq0$.
        \item Revisitation does not occur, i.e., $\Tilde{h}_{vw}^{\text{REV}}=1$:\\
            $\hat{a}_{vw}\leq \sum_{e_{pq}^{(i)}\in\mathcal{E}^{\text{P }}}D_{pqi}^{\text{RTT}}(1-\Tilde{a}_{vw})$,\\
            $\hat{a}_{vw}\leq \sum_{e_{pq}^{(i)}\in\mathcal{E}^{\text{P }}}D_{pqi}^{\text{RTT}}\Tilde{u}_{vw}^{pqi}-D_{vw}^{\text{REQ}} + \sum_{e_{pq}^{(i)}\in\mathcal{E}^{\text{P }}}D_{pqi}^{\text{RTT}}\Tilde{a}_{vw}$\\
            In this case, $\Tilde{a}_{vw}$ will be determined based on the value of $D_{pqi}^{\text{RTT}}$, $D_{vw}^{\text{REQ}}$, and $\Tilde{u}_{vw}^{pqi}$ with always only one scenario being feasible.
    \end{itemize}
    
    } {\color{black}Therefore, in all cases, the four constraints mentioned above can accurately replace the term in (\ref{eq33}).} As a result, by relaxing $\Tilde{a}_{vw}$, (\ref{eq38}) and the objective become an affine set and function, respectively.
    
    The next non-convex term in 5G-INS is the ceiling function $\lceil\cdot\rceil$ used in (\ref{eq13}), (\ref{eq17}), (\ref{eq29}), and (\ref{eq38}). {\color{black}In (\ref{eq13}), $h_p^{\text{L0}}$ can be accurately expressed by a set of piecewise linear functions with segments defined in unit intervals within the range $[0,\Xi_{p,1}^{\max}]$, i.e., $[0,1]$, $[1,2]$, $\cdots,$ $[\Xi_{p,1}^{\max}-1,\Xi_{p,1}^{\max}]$. However, we can also approximate it by its linear upper bound, $\cdot + 1$, which will result in negligible approximation since, in the worst case, it will result in at most one VM of difference for a server. It also ensures that all constraints remain within their feasible range and provides additional flexibility for the hypervisor, host OS, and OpenStack.
    
    Other occurrences of $\lceil\cdot\rceil$ are associated with the terms involving L1 SRs. We first change the definition of $B^{\text{REQ}}_{vw}$ and $K_w^{\text{REQ}}$ to represent the bandwidth consumed by the allocated number of subcarriers and the number of allocated PRBs in full radio frames instead of the initial requested bandwidth and PRBs. For L1 SRs, since PRB allocation and bandwidth allocation occur in full radio frames and dedicated subcarriers, arbitrary QoS violation is not allowed. In addition, the deployment cost will discourage overprovisioning. Therefore, $ky_w^{q,\sigma,k}/N^{\text{FRM}}=\lceil ky_w^{q,\sigma,k}/N^{\text{FRM}}\rceil$ and $b_{vw}^{pqi,l,\lambda}/B^{\text{WSC}}=\lceil b_{vw}^{pqi,l,\lambda}/B^{\text{WSC}}\rceil$. Consequently, the relaxed 5G-INS problem can be accurately transformed into a linear program.}
    \begin{figure}
	\centering
	\includegraphics[width=0.85\linewidth]{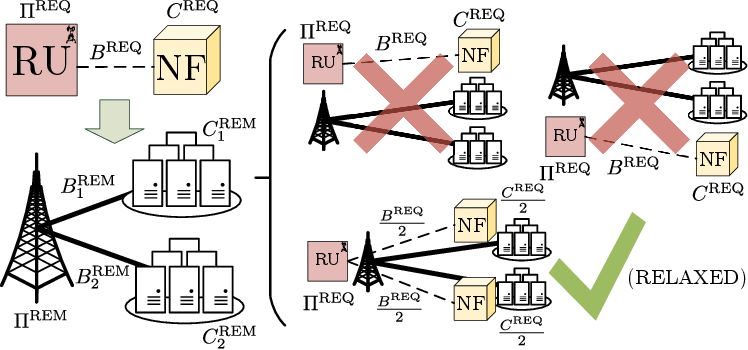}
	\caption{\textcolor{black}{The NS scenario used in Theorem 3. Remaining MIPS capacities, i.e., $C^{\text{REM}}_1$ and $C^{\text{REM}}_2$, are insufficient to satisfy the MIPS requirement, i.e., $C^{\text{REQ}}$, individually, making the two mappings shown in the top-right corner infeasible. However, \(C^{\text{REM}}_1+C^{\text{REM}}_2\geq C^{\text{REQ}}\). With each server hosting $C^{\text{REQ}}/2$, the relaxed mapping in the bottom-right corner becomes feasible.
    }}\label{th3}
    \end{figure}
    \section{Proof of Theorem 3}\label{appc}
    {\color{black}To prove the theorem, it suffices to demonstrate a scenario where the relaxed 5G-INS problem is feasible and the original problem is not. Consider the small example in Fig. \ref{th3}, where all pre-embedded SRs and the arrived SR have identical isolation levels and are HP, thereby no QoS violations. Let \( C^{\text{REQ}} \), \( B^{\text{REQ}} \), and \( \Pi^{\text{REQ}} \) represent the required MIPS, bandwidth, and radio resources, respectively, and \( C^{\text{REM}}_1 \), \( C^{\text{REM}}_2 \), \( B^{\text{REM}}_1 \), \( B^{\text{REM}}_2 \), and \( \Pi^{\text{REM}} \) denote the remaining capacities. Assume 1) $\Pi^{\text{REM}} > \Pi^{\text{REQ}}$, 2) $ B^{\text{REM}}_1, B^{\text{REM}}_2 > B^{\text{REQ}}$, and $C^{\text{REQ}}/2 < C^{\text{REM}}_1, C^{\text{REM}}_2 < C^{\text{REQ}}$. In the original 5G-INS problem, \( C^{\text{REM}}_1 \) and \( C^{\text{REM}}_2 \) are insufficient to host \( C^{\text{REQ}} \) individually, making it infeasible. In the relaxed 5G-INS, \( C^{\text{REQ}} \) can be split between \( C^{\text{REM}}_1 \) and \( C^{\text{REM}}_2 \), making it feasible. This completes the proof.}
\end{appendices}

\bibliographystyle{IEEEtran}
\bibliography{references}

\end{document}